\documentclass[aps,10pt,twocolumn,tightenlines]{revtex4}
\usepackage{setspace}
\usepackage{amsmath}
\usepackage{epsfig}
\newcommand{\be}{\begin{equation}}
\newcommand{\ee}{\end{equation}}
\newcommand{\ba}{\begin{eqnarray}}
\newcommand{\ea}{\end{eqnarray}}

\begin{document}

\title{Radio Cherenkov signals from the Moon: neutrinos and cosmic rays}

\author{Yu Seon Jeong and Mary Hall Reno}
\affiliation{Department of Physics 
and Astronomy, University of Iowa, Iowa City, IA 52242}
\author{Ina Sarcevic}
\affiliation{Department of Physics, University of Arizona, Tucson, AZ 85721 \\
Department of Astronomy and Steward Observatory,
University of Arizona, Tucson, AZ 85721}
\begin{abstract}

Neutrino production of radio Cherenkov signals in the Moon is the object of radio telescope observations.
Depending on the energy range and detection parameters, the dominant contribution to the
neutrino signal may come from interactions of the neutrino on the Moon facing the telescope, rather than
neutrinos that have traversed a portion of the Moon.  Using the approximate analytic expression of the
effective lunar aperture 
from a recent paper by Gayley, Mutel and Jaeger, 
we evaluate the background
from cosmic ray interactions in the lunar regolith.  We also consider the modifications to the
effective lunar aperture from generic non-standard model neutrino interactions.
A background to neutrino signals are radio Cherenkov signals from cosmic ray interactions.
For cosmogenic neutrino fluxes, neutrino signals will be difficult to observe 
because of low neutrino flux at the high energy end and 
 large cosmic ray background in the lower energy range considered here.  
We show that lunar radio detection of neutrino interactions 
is best suited to constrain or measure neutrinos from astrophysical sources 
and probe non-standard neutrino-nucleon interactions such as microscopic 
black hole production.  
\end{abstract}
\pacs{xx}

\maketitle

\section{Introduction}

Ultrahigh energy neutrinos may originate from astrophysical sources, 
from exotic sources such as ultramassive particles which decay and from cosmic ray interactions
with the background radiation \cite{review}.  Because cosmic rays have been observed up to 
energies of $10^{19}$ eV, 
high energy neutrino flux from cosmic ray 
interactions with photons producing charged pions \cite{gzk} which decay into 
neutrinos \cite{bz} is a ``garanteed'' neutrino flux. 
Deatiled flux predictions of
these final
state neutrinos, called GZK-neutrinos or cosmogenic neutrinos, still have
theoretical uncertainties associated with the composition and injection
spectrum of the highest energy
cosmic rays, and the photon spectum now and at earlier epochs, as 
discussed in, e.g., Refs.
 \cite{hs,pj,ess,hts,oka,olinto}. Observations
of neutrino signals will be an important piece of the high energy astrophysics and
cosmological picture. 

In contrast to photons, neutrinos have weak interaction cross sections 
\cite{gqrs,renonu,jeong1,jgr,ctw} 
so neutrino fluxes are not attenuated over
cosmic distances.
However, one needs detectors sensitive to 
many targets for neutrino interactions since the 
interaction probability is low.
There are already a number of effort to observe these highest
energy neutrinos, including signals from the air showers they would produce \cite{augernu}, and
from the particles and radiation they produce when they interact in matter
 \cite{icecubeuhenu,anita,Anita2,hankins,gorham,Jaeger,westerbork,nuMoon,lunaska,LOFAR}. All observational
efforts require large volumes.

Among these observational efforts are neutrino
induced events on the Moon, where the Moon is the target, and the signal is the radio Cherenkov emission
\cite{hankins,gorham,Jaeger,westerbork,nuMoon,lunaska,LOFAR}.  
Neutrinos, when they interact with nucleons and nuclei, generate hadronic showers with shower 
energies of $E_{\rm shr}\simeq 0.2 E_\nu$.
An electron charge excess is produced, and this group of electrons moves 
faster than the speed of light in the lunar regolith, hence the Cherenkov signal
\cite{askaryan}.

Evaluation of the signal depends on the cosmogenic neutrino flux,
the ultrahigh energy neutrino cross section, 
the radio signal production, attenuation and refraction at the surface and detection parameters. 
An approximate expression for the effective aperture for neutrino
induced radio Cherenkov signals from the Moon has been
developed by Gayley, Mutel and Jaeger(GMJ) in Ref. \cite{gayley}. Ultrahigh energy neutrinos
are incident isotropically on the Moon to a good approximation.
In analogy with terrestrial observations, e.g.,  at the IceCube neutrino observatory
\cite{icecube}, neutrinos that are incident on the ``backside'' of the
Moon and traverse a portion of the Moon before interacting are called
``upward'' neutrinos. Neutrinos which interact on the surface of
the Moon facing the Earth are denoted ``downward'' neutrinos.

Gayley et al. find that, depending on the energy and cross section, the event 
rate from neutrinos incident on the surface we see (downward
neutrinos) sometimes dominate over the event 
rate from neutrinos incident on the surface of the Moon not visible from Earth 
(upward neutrinos) \cite{gayley}. 
This
may seem counter-intuitive, however, between the various angles of 
incidence, the Cherenkov angle and the angular spread of the
Cherenkov cone, 
angles of refraction and angle characterizing the lunar surface roughness, 
the downward neutrinos can contribute appreciably to the signal.

\begin{figure}[t]
\begin{center}
\includegraphics[angle=270,width=0.45\textwidth]{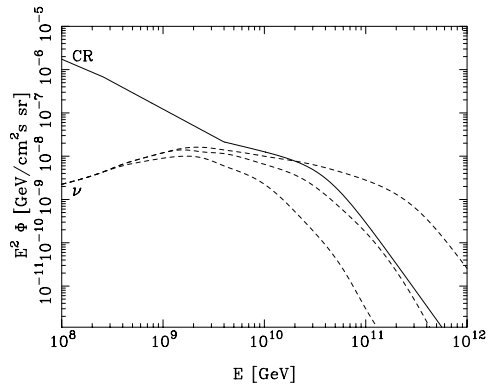}
\end{center}
\caption{The flux of cosmic rays (solid) { with the parametrization from Ref. \cite{crflux}} and three models for the flux of cosmogenic neutrinos (dashed) from Ref. \cite{olinto} with the assumption of proton primaries.}
\label{fig:flux}
\end{figure}

Cosmic ray interactions in the lunar regolith produce hadronic showers
as well. 
These cosmic ray induced hadronic showers also induce a charge excess and a 
related radio Cherenkov signal which can 
be used to measure or constrain the
ultrahigh energy cosmic ray flux \cite{scholten}. The cosmic ray flux is quite large at low energy, but it falls off rapidly with increasing energy. The flux of cosmic rays (solid line) and three models for the flux of cosmogenic neutrinos from Ref. \cite{olinto} (dashed lines) are shown in Fig. \ref{fig:flux}. The three models for
the cosmogenic neutrinos are a subset of predictions for the cosmogenic neutrino flux. These three come from the initial assumption of cosmic rays which are primarily protons, in which the sources are assumed to have maximum acceleration energies of $E_{p,max}=10^{11}$ GeV, $10^{11.5}$ GeV and $10^{12}$ GeV. Details of the model, including injection spectra, assumptions about the star formation rate and the correlated UHE cosmic ray sources and other inputs to the evaluation are described in Ref. \cite{olinto}. The cosmic ray flux shown uses the 
parametrization of the Auger Collaboration in Ref. \cite{crflux}.

Given the cosmic ray flux \cite{crflux} and its production of radio Cherenkov signals similar to the corresponding neutrino
Cherenkov signals \cite{scholten}, we evaluate the cosmic ray induced signals.  The considerations
for cosmic rays, as compared to neutrinos, are somewhat different.
Cosmic ray fluxes are attenuated in
the Earth's atmosphere, however, there is no lunar atmosphere.
Within the lunar regolith, the cosmic ray interaction length is
small compared to any other characteristic distance scale.
There will not be any ``upward'' cosmic ray induced signals, but there
is the potential for ``downward'' cosmic rays to produce radio
Cherenkov signals. Cosmic ray flux attenuation in the lunar regolith is, of course, a large effect, { resulting
in cosmic ray signals being produced very close to the lunar surface. In Ref. \cite{scholten}, ter Veen {\it et al.}
have determined that the radio Cherenkov signal far from the lunar surface is essentially the same as one originating
deeper in the regolith (apart from attenuation).}

In this paper, we look at the relative importance of cosmic ray
and neutrino induced radio Cherenkov signals from the Moon assuming standard model cross sections and the
neutrino and cosmic ray fluxes in Fig. 1. 
We use the approximate expression from Ref. \cite{gayley} for the event rates, 
and we modify this analytic result to account for strong attenuation where applicable. We also consider
non-standard model neutrino-nucleon cross sections and generic neutrino fluxes from astrophysical sources.  

There have been a number of discussions of the cross section dependence
of various neutrino induced signals 
\cite{hussain,kusenko,acceptances}. In the next section, we review the neutrino cross section and the resulting
dependence of the event rate on the cross section. 
In Section III, we evaluate the event rates using the standard model 
neutrino nucleon cross section and for some
rescaled cross sections, all using the approximate analytic GMJ
expression of Ref. \cite{gayley}. The dependence of the event rate on {
detection characteristics, namely}
the radio frequency and minimum detectable electric field, is shown.

In Section IV, we show how the approximate analytic expression is
modified to account for cosmic ray flux attenuation on short distance scales in the lunar regolith. We find that the event rate is independent of cosmic ray cross section, as long  as the cosmic ray interaction length is small compared to the photon attenuation
length. We compare the
rates of cosmic ray and neutrino induced radio signals using the fluxes shown in Fig. 1
{and the standard model neutrino nucleon cross section}. We show
in Section V
how the inclusion of mini-black hole production, as an example of 
a neutrino-nucleon cross section enhancement,  and how alternative neutrino spectra affect the predicted event rates. 

Our conclusions appear in Section VI. We find that while lowering the minimum electric field detectable
by a radio telescope array would help increase the number of neutrino events, since it effectively lowers the
neutrino energy threshold for detection, it also increases the number of cosmic ray events. Cosmogenic neutrino fluxes on the scale
of those presented in Ref. \cite{olinto} will be difficult to observe on the one hand because of low fluxes (at the high energy end)
or because of the cosmic ray background (in the lower energy range considered here). Lunar radio detection of neutrino interactions 
is best suited to constrain or measure neutrino sources other than the cosmogenic sources and non-standard neutrino nucleon
cross section enhancements.

\section{Neutrino cross sections and effective solid angle}

The dependence of neutrino induced events on the neutrino cross section has been 
the subject of much discussion, e.g., in Refs. \cite{hussain,kusenko,acceptances}.
One feature is that the probability of interaction is proportional to the 
neutrino nucleon cross section, however, the neutrino flux attenuation is also
affected by the cross section.

Experimentally, the neutrino nucleon cross section has been directly measured for
$E_\nu<450$ GeV \cite{pdg}. A related measurement, the charged current interaction
cross section for electrons in $ep$ scattering at HERA translates to a neutrino
cross section with $E_\nu=27 $ TeV incident on a proton at rest \cite{hera}.
The moment transfer relevant to UHE neutrino scattering is 
characterized by the W-boson mass $M_W$. At $Q^2\sim
M_W^2$, the structure functions have been measured in the Bjorken $x$ regime
of $x$ larger than a few times $10^{-3}$. At the Large Hadron Collider, one
expects measurements of the structure functions for $x>10^{-5}$ for similar
values of $Q^2$. The reach in $x$ at the Large Hadron 
Collider (LHC) extends to an
equivalent neutrino energy $E_\nu\sim 10^8$ GeV, using the approximate correspondence that
$2M_N E_\nu \sim M_W^2\simeq \langle Q^2\rangle$ \cite{gqrs}.
We consider here neutrino energies $E_\nu\geq 10^8$ GeV, above
kinematic regions probed even by the LHC experiments.

\begin{figure}[h]
\begin{center}
\includegraphics[angle=270,width=0.45\textwidth]{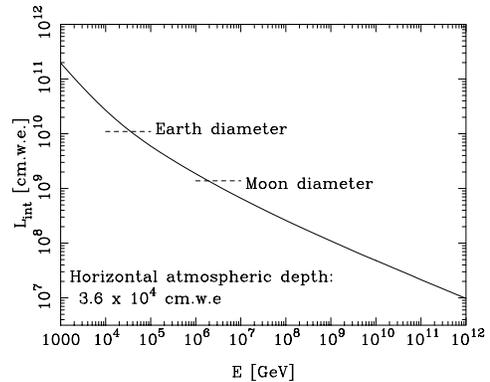}
\end{center}
\caption{The neutrino interaction length (in centimeters water equivalent distance)
 as a function of energy, with indications
of the diameter of the Earth and Moon { using the cross sections from Ref. \cite{jeong1}}.}
\label{fig:lint}
\end{figure}

The neutrino cross section in the standard model, with a power law
extrapolation of the structure functions at low $x$, has an approximate
power law energy dependence\cite{gqrs,renonu} $\sigma_{\nu N}^{\rm tot}\simeq 1.57\times 10^{-35}
\ {\rm cm}^2 (E_\nu/{\rm GeV})^{0.33}$.
The neutrino interaction length is defined to be
\begin{equation}
L_{\nu} \equiv \frac{1}{\sigma_{\nu N}^{\rm tot} N_A}\ .
\end{equation}
This has the dimensions of g/cm$^2,$ or equivalently centimeters of
water equivalent distance (cmwe), which can be converted to a distance
by dividing by density. 
Rather than use a power law form, we use the charged current
cross sections of Ref. \cite{jeong1}, with the addition of the top quark
contribution (on the order of a 30\% correction at the highest
energy \cite{jeong1,jgr}). The charged current cross section,
for high energies,
is multiplied by  an approximately energy independent
constant of 1.43 to obtain the total neutrino-nucleon cross section \cite{ctw}.
The standard model neutrino interaction length
is shown in Fig. \ref{fig:lint}. For an incident neutrino energy
of $10^3$ GeV, the tau neutrino cross section is only $\sim 5\%$ less than the
muon neutrino cross sections \cite{jeong2}. 
We will be considering higher energies where
we can use the same cross sections for all three neutrino flavors.

The importance of neutrino attenuation in the Moon or Earth can be seen with a comparison of the interaction length $L_{\nu}$
as a function of neutrino energy for the standard model cross section with
the Earth and Moon diameters. The Earth
diameter equals the neutrino interaction length for $E_\nu$ larger than
a few tens of TeV. 
The Moon's diameter (in cmwe) equals the neutrino
interaction length for an incident neutrino energy of a few times
$10^6$ GeV. Our focus will be on cosmogenic neutrinos at energies much higher than $10^6$ GeV, so attenuation effects will be important in the
evaluation of interaction rates, as we discuss in detail below. 
We note that the { Earth's}
horizontal atmospheric depth is small on the scale of the neutrino interaction length, even at $E_\nu=10^{12}$ GeV.

The cross section dependence of event rates is specific to the signal and the 
possibility of neutrino regeneration
through neutrino neutral current interactions and for tau neutrinos,
through neutrino production and decay \cite{hussain,kusenko,dutta}.
The neutral current neutrino regeneration can be approximately included via an
effective cross section
\begin{equation}
\sigma\equiv\sigma_{\nu N}^{\rm eff} = \kappa \sigma_{\nu N}^{\rm tot},
\end{equation}
where we have
taken $\sigma$ to scale with energy as $\sigma_{\nu N}^{\rm tot}$.
In principle, $\kappa$ depends on neutrino energy, but in practice at high energy, the differential cross section for neutrino neutral current interactions has an approximate scaling with neutrino energy the same way as the total cross section. We neglect the $\nu_\tau\to \tau\to\nu_\tau$ regeneration, which is
typically not very important for steeply falling fluxes \cite{dutta}.

We can estimate $\kappa$ in the standard model with some approximations. This is also outlined in the Appendix of Ref. \cite{gayley}. The neutrino flux 
as a function of column depth $X$ is
\begin{eqnarray}
\frac{d\Phi_\nu(E,X(\theta ))}{dX} &=& -\frac{\Phi_\nu (E,X(\theta ))}{L_\nu}\\
\nonumber
&+
&\int_{E_\nu}^\infty
dE'\Phi_\nu(E',X(\theta ))\frac{d\sigma_{NC}(E',E)}{dE}
\label{eq:nutransport}
\end{eqnarray}
In the standard model, the neutral current cross section is approximately $r_{NC}=0.3$ of the total cross section. Using
\begin{equation}
\frac{d\sigma_{NC}(E',E)}{dE}\simeq r_{NC}\sigma_{\nu N}^{\rm tot}(E')
\delta (E-(1-\langle y\rangle)E')
\end{equation}
where $y=(E'-E)/E'$ is the neutrino inelasticity, relating the change in neutrino energy when it interacts, normalized to the initial neutrino
energy. At high energies, $\langle y\rangle \simeq 0.2$. For a power
law spectrum $\Phi_\nu(E,X(\theta ))\sim E^{-\gamma}$ and for a
neutrino cross section which scales with energy as
$\sigma_{\nu N}^{\rm tot}\sim E^\delta$, Eq. (\ref{eq:nutransport})
can be written as
\begin{eqnarray}
\nonumber
\frac{d\Phi_\nu(E,X(\theta ))}{dX} &=& -\Bigl(1- r_{NC} (1-\langle y\rangle )^{\gamma -1 - \delta}\Bigr)
\frac{\Phi_\nu(E,X(\theta ))}{L_\nu}\\
&=& - \kappa \frac{\Phi_\nu(E,X(\theta ))}{L_\nu}\ .
\end{eqnarray}
When $\delta = 0.3$, as is approximately the case
for the standard model, with $r_{NC} = 0.3$, $\kappa \simeq 0.74$
for a spectral index $\gamma = 2$. The value of $\kappa$ is not very
sensitive to the spectral index. It increases to
$\kappa\simeq 0.84$ when $\gamma=4$. We use $\kappa = 0.84$ in our evaluation
below, and we define the attenuation distance $\lambda$ to
include the regeneration effect by 
\begin{equation}
\lambda\equiv 1/\sigma N_A
= L_{\nu}/\kappa\ . 
\end{equation}

We review here the scaling
of upward event rates including attenuation as a function of the cross section.
Schematically, event rates $\Gamma$ for a detector of cross sectional area
$\cal{A}$ are given by
\begin{eqnarray}
\nonumber
\Gamma &=&\int dE_\nu \, d\Omega_\nu \, d\hat{\cal A}
\cdot \hat{n}(\theta_\nu) \,  dr {\cal P}(E_\nu,\theta_\nu,r)
\\
\label{eq:factorize}
&\times & \Phi_\nu(E_\nu, X(\theta_\nu))
\end{eqnarray}
where $\Phi_\nu(E_\nu,X)$ is the neutrino flux in units of neutrinos/(cm$^2$s sr GeV),
and $dr\, {\cal P}(E_\nu,\theta_\nu)$ is the probability the neutrino of energy $E_\nu$ produces a
signal in the interval $dr$. The angle $\theta_\nu$ is the incident angle of the neutrino flux with respect to vector normal to the cross sectional area of the detector.
The probability to produce a signal depends
linearly on the neutrino cross section for the specific signal ($\sigma_s$),
so considering short distances $L=\int dr$, we can write
$dr\, {\cal P}(E_\nu,\theta_\nu)=dr \sigma_s N_A\rho$. 
The effective volume of the detector is $V$, where $dV = drd{\cal A}$. 
We consider detection near the surface, where for upward neutrino
fluxes, the depth of the
detector is negligible.

We start with a configuration of a detector near the surface which is approximately isotropic: where $\hat{\cal A}\cdot\hat{n}$ is independent
of the neutrino direction and where the pathlength of the neutrino in the detector of size $V=L^3$ is approximately independent of incident neutrino direction:
$${\cal P}^{\rm iso}(E_\nu,\theta_\nu,L)\simeq {\cal P}^{\rm iso}(E_\nu,0,L)\ .$$
The neutrino flux accounting for 
attenuation, for an { upward} neutrino traversing a sphere and emerging with an
angle $\theta_\nu$ with respect to the normal to the surface is approximately
\begin{equation}
\label{eq:shadow}
\Phi_\nu (E_\nu,X(\theta_\nu ))\simeq e^{-2R\cos\theta_\nu/\lambda}
\Phi_\nu (E_\nu,0)
\end{equation}
for a flux incident on the Earth or Moon with column depth
of the diameter of $2R$. For the Moon, $2R = d_M\rho_{M}$ for the diameter
of the Moon $d_M= 3,480 $ km and average lunar density $\rho_{M}=3.34$ g/cm$^3$. 
Here, $\theta_\nu$ is the zenith angle of the incident neutrino flux. We assume
that the neutrino flux is isotropically incident on a spherical body
(the Earth or Moon). 

\begin{figure}[h]
\begin{center}
\includegraphics[angle=270,width=0.45\textwidth]{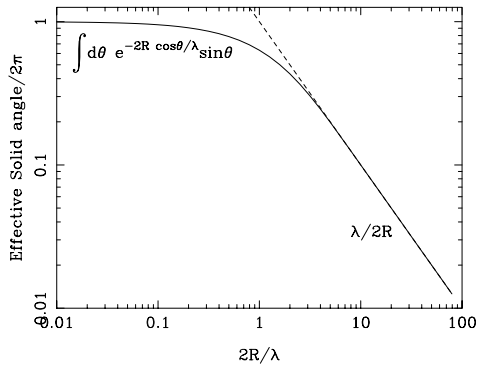}
\end{center}
\caption{The effective solid angle divided by $2\pi$ for an underground detector with an isotropic effective area, as a function of the ratio of the diameter (in water
equivalent distance) to
the attenuation distance (solid line). The dashed line shows $(\lambda/2R)$.}
\label{fig:effang1}
\end{figure}

By factorizing the depth dependent neutrino flux as in 
Eq. (\ref{eq:shadow}), the integrand for an isotropic
incident flux has a factor which includes the effective solid 
angle $\Omega_{\rm eff}$ where
\begin{eqnarray}
\nonumber
\frac{\Omega_{\rm eff}^{\rm iso}}{2\pi}  
&=& 
 \int_0^{\pi/2} d\theta_\nu\, \sin\theta_\nu 
e^{-2R\cos\theta_\nu/\lambda}\\
& = & \frac{\lambda}{2R} \Bigl(1-\exp(-2R/{\lambda})\Bigr) .
\label{eq:omegaiso}
\end{eqnarray}
The angular integral for a fixed energy (fixed $\lambda$) depends only on $2R/\lambda$. This is shown in Fig. \ref{fig:effang1} where
the dashed line shows the quantity $\lambda/2R$. 
The dashed line is the scaling
behavior of the integral for small interaction lengths compared to the
diameter of the Earth or Moon ($\lambda/2R<1/2$).

\begin{figure}[h]
\begin{center}
\includegraphics[width=0.40\textwidth]{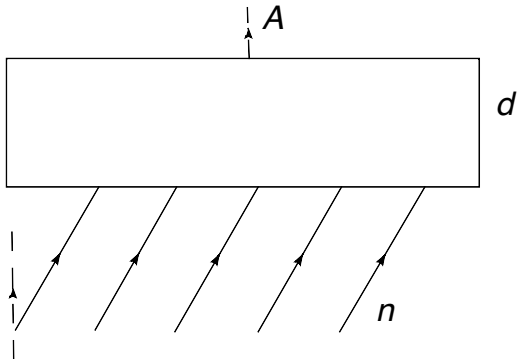}
\end{center}
\caption{A detector of area $\cal A$ and depth $d$, with neutrinos incident
in a directions characterized by $\hat{n}$ which makes an angle $\theta_\nu$
to the normal to the surface $\hat{\cal A}$.}
\label{fig:rectdetect}
\end{figure}

For detectors which are not isotropic, similar result is found. Consider
the a detector schematically shown in Fig. \ref{fig:rectdetect}, 
where the area is ${\cal A}=A$ and the detector 
thickness is $d$, the quantity $\hat{\cal A}\cdot\hat{n}={\cal A}\cos\theta_\nu$
appears in the integral, but now $\int dr {\cal P}(E_\nu,\theta_\nu,r)=\sigma_s N_A\rho
d/\cos\theta_\nu$, assuming no neutrino 
attenuation through the depth of the detector. The combined angular dependence
makes the integrand the same as for an isotropic detector, so $\Omega^{\rm iso}_{\rm eff}$ is the same for any isotropic upward neutrino flux incident
on an underground detector. Eq. (\ref{eq:omegaiso}) shows that roughly, for a given $\lambda/R$, the zenith angles which contribute 
range between $\theta = 90^\circ$ and $\theta = \cos^{-1}(\lambda/2R)$ when $2R\gg
\lambda$.

Combined with the neutrino interaction probability and
the effective area ${\cal A}={\cal A}_0$, the cross section dependence
of the effective solid angle, for small interaction lengths relative to the
column depth $2R$ is
\begin{equation}
\label{eq:scaling}
\int dr \Omega_{\rm eff}^{\rm iso}P(E_\nu,0,r) {\cal A} \sim {\cal A}_0 \frac{\sigma_s}{\sigma}
\frac{d\rho}{2R}\  
\end{equation}
for isotropic incident fluxes.
This agrees with the discussion of, e.g., Ref. \cite{hussain}, where it is noted that for large neutrino cross sections, the effective solid angle
is reduced in just the proportion to the increase in event rate
in the detection region in their discussion of upward shower events.

The slab configuration of Fig. \ref{fig:rectdetect} is relevant to neutrino
induced radio Cherenkov signals.
As we discuss below, radio signals are produced in the a thin layer of the lunar
regolith on the surface facing the Earth. The depth of layer
is characterized by the attenuation length of the radio signal, $L_\gamma$. Additional
factors are required to account for the radio signal production and refraction
at the surface, but schematically, the upward signal will be independent of
the neutrino cross section at high energies, and it is proportional to $L_\gamma$.

\section{Lunar Radio Cherenkov signals from neutrinos}

To evaluate the neutrino event rate, we look at the effective 
aperture { of the Moon} evaluated by
Gayley, Mutel and Jaeger in Ref. \cite{gayley}.
The effective aperture $A_M(E_\nu)$ combines the neutrino interaction probability, effective area (${\cal A}$) and effective solid angle,
\begin{equation}
\label{eq:evtrate}
\Gamma = \int dE_\nu A_M(E_\nu)\Phi_\nu (E_\nu,X)\ .
\end{equation}
Eq. (\ref{eq:evtrate})  also accounts for the radio Cherenkov signal production,
attenuation and refractions at the lunar surface. $A_M(E)$ has units of
sr$\cdot$km$^2$.
GMJ\cite{gayley} have shown that it is possible to
parametrize the effective aperture for a lunar radio Cerenkov signal
with the analytic form
\begin{eqnarray}
\nonumber
A_M(E) &= &A_0 \frac{(n_r^2-1)}{8n_r}\frac{L_\gamma}{L_\nu} f_0^3\Delta_0\\
\nonumber
&\times & 
(\Psi_{ds}+\Psi_{dr} +\Psi_u) \ \\
&=& A_{ds}+A_{dr}+A_u \nonumber \\
&\equiv & A_0 P(E) 
\ .
\label{eq:aperture}
\end{eqnarray}
In this equation for the Moon, the index of refraction is $n_r=1.73$ 
and the maximum lunar aperture is $A_0=4\pi (\pi R_M^2)$ for lunar
radius $R_M=1,740$ km. The photon interaction length (in g/cm$^2$ units) is $L_\gamma = 9 \ {\rm m}\times \rho/(\nu / {\rm GHz})$
with a lunar regolith density of $\rho=1.7$ g/cm$^3$.
We discuss our neutrino interaction length in detail below, but this factor of $L_\nu^{-1}$ carries the factor of $\sigma_s$ in the probability for the neutrino
to produce a signal. It is convenient to scale out a factor of $A_0$ and look at
the energy dependent function $P(E)$.

{The parameter $f_0$ is the ratio of the thickness of the Cherenkov cone at the electric
field threshold $\varepsilon_{\rm min}$ to the full thickness of the Cherenkov cone ($2\Delta_0$)},
\begin{equation}
f_0 = \sqrt{\ln\Biggl( \frac{0.6\varepsilon_0}{\varepsilon_{\rm  min}}\Biggr)}
\end{equation}
{ where already the requirement that the electric field of the signal at the Earth is larger than
the electric field threshold of the detector has been enforced.}
The quantity $\varepsilon_0$
depends on the distance to the moon $d=3.84\times 10^5$ km, the energy of the neutrino induced shower
$E_{shr}$
and the radio frequency of the radiation $\nu$ is
\begin{eqnarray}
\nonumber
\varepsilon_0 &=& 0.0845\ \frac{\rm V}{\rm m\, MHz}\Biggl[ \frac{d}{\rm m}\Biggr]^{-1}
\Biggl[\frac{E_{\rm shr}}{\rm EeV}\Biggr]
\Biggl[\frac{\nu}{\rm GHz}\Biggr]\\
&\times & 
\Biggl[ 1+\Biggl( \frac{\nu}{2.32\ {\rm GHZ}}\Biggr)^{1.23}\Biggr]^{-1}
\end{eqnarray}
where $E_{\rm shr}\simeq 0.2 E_\nu$ is the approximate hadronic shower energy for neutrino interactions with nucleons in the lunar
regolith. The quantity $\Delta_0$ is the Cherenkov cone half width,
\begin{equation}
\Delta_0=0.05 \Biggl[\frac{\rm GHz}{\nu}\Biggr]\Biggl[
1+0.075 \log_{10} \Biggl( \frac{E_{\rm shr}}{10^{10}\ {\rm GeV}}\Biggr)\Biggl]^{-1}\ .
\end{equation}
{ The electric field threshold $\varepsilon_{\rm min}$ of the detector depends on
the collection area of the telescopes, the band width and other features of the
specific telescope or array of telescopes \cite{gayley}. One example is
Project RESUN \cite{Jaeger} using the radio Expanded Very Large Array
where $\varepsilon_{\rm min}\sim 10^{-8}$ V/m/MHz at 1.4 GHz.}

There are three terms in eq. (\ref{eq:aperture}) 
representing the angular aperture fractions for downward neutrinos on a
smooth surface (ds), downward neutrinos on a rough surface (dr) and upward
neutrinos (u). They are respectively,
\begin{eqnarray}
\nonumber
\Psi_{ds} &=& f_0\Delta_0\\
\nonumber
\Psi_{dr} &=& \frac{16}{3\pi^{3/2}}\sigma_0 = 0.96\sqrt{2}
\tan^{-1}(0.14\nu^{0.22})\\
\Psi_u & = & \frac{16}{3}\frac{\lambda}{2R_M\rho_M}=\frac{16}{3}\frac{L_\nu}{2R_M\rho_M\kappa}\ .
\end{eqnarray}
These terms were derived in the approximation that the neutrino interactions occur near the lunar surface, and that the spread of the Cherenkov cone is small, as is the incident neutrino direction relative to the horizontal to the lunar surface. This last approximation is valid as long as the neutrino interaction length is small compared to the lunar diameter.

Qualitatively, the prefactor accounts for the incident angles of the neutrinos, the cross sectional area of the moon, the interaction probability in the outer layer of the lunar regolith
$$P_{\rm eff} \sim \frac{L_\gamma}{L_\nu} \sin\theta_c f_0^2\ ,$$
the integral over the width of the Cherenkov cone  giving $\sin\theta_c f_0 \Delta_0$ and the refraction effect on the solid angle subtended by the emerging radiation from inside the Moon ($n_r$). { We recall that $\sin^2\theta_c= (n_r^2-1)/n_r^2$.} 

For the upward effective aperture, the factor $\Psi_u$ accounts for the reduced effective solid angle due to neutrino attenuation in the Moon. In fact,
\begin{equation}
\label{eq:psiu}
\Psi_u  =  \frac{16}{3}\frac{\Omega_{\rm eff}^{\rm iso}}{2
\pi}\ .
\end{equation}
If the neutrino nucleon cross section is smaller than the standard model cross section, or at lower energies, the scaling of $\Omega_{\rm eff}^{\rm iso}$ with $(\lambda/2 R)$ is modified, so we make the substitution of eq. (\ref{eq:psiu})
in the event rate to allow for lower cross sections (larger interaction lengths).

For neutrinos that are incident downward, some radio signal will emerge, namely the portion of the solid angle equal to the thickness of the Cherenkov cone. 
Without surface roughness, this is the only contributor to the radio signal, however, surface roughness permits the radio signal that would otherwise be lost to emerge. The approximate analytic result of Gayley et al. has the
$\Psi_{dr}$ contribution proportional to $\sigma_0=\sqrt{2}\tan^{-1}(0.14\nu^{0.22})$, the surface roughness parameter in terms
of the radio frequency $\nu$ in GHz. For $\nu=1.5$ GHz, $\sigma_0=12.3^\circ=0.21$
rad.

\begin{figure}[t]
\begin{center}
\includegraphics[angle=270,width=0.45\textwidth]{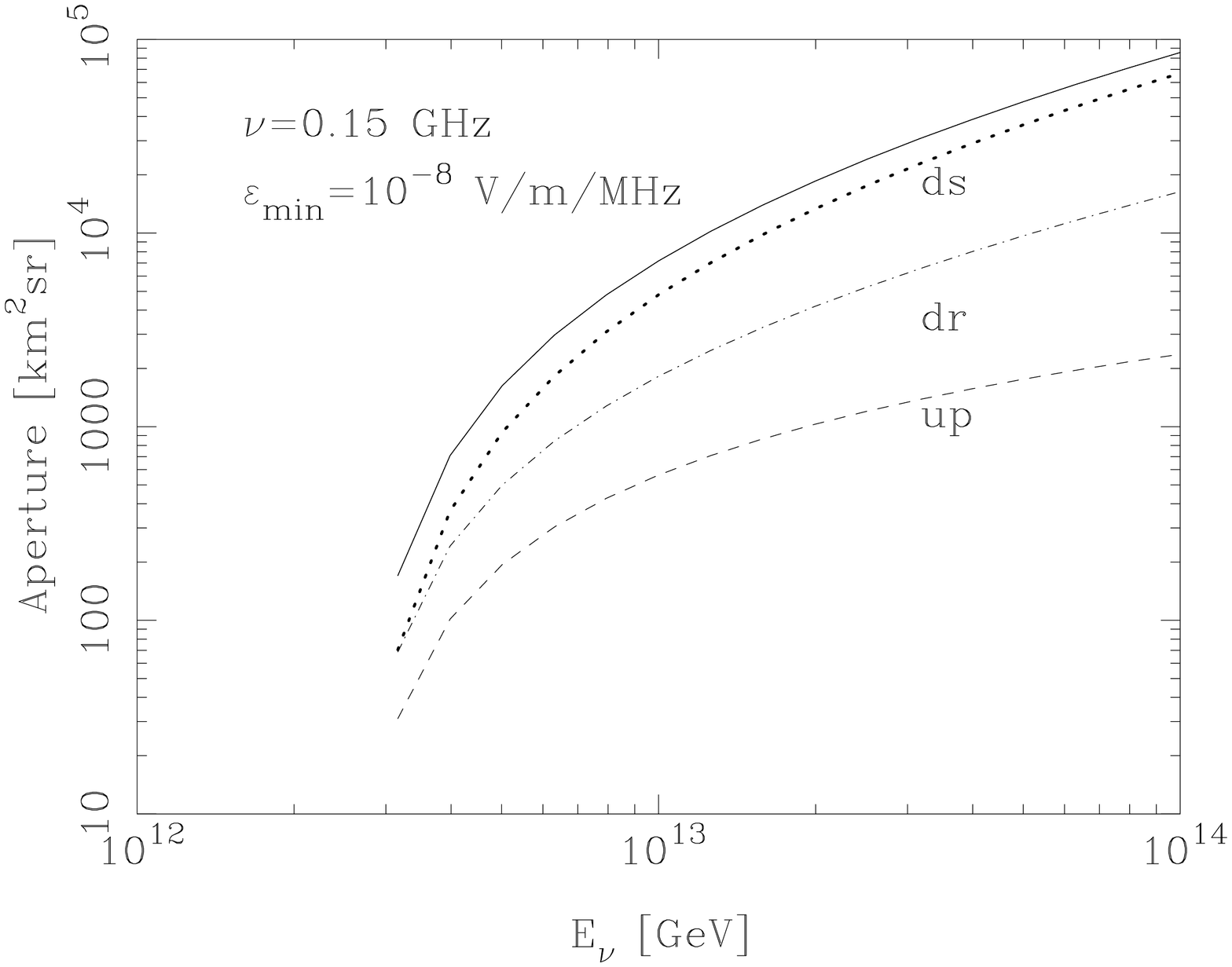}
\includegraphics[angle=270,width=0.45\textwidth]{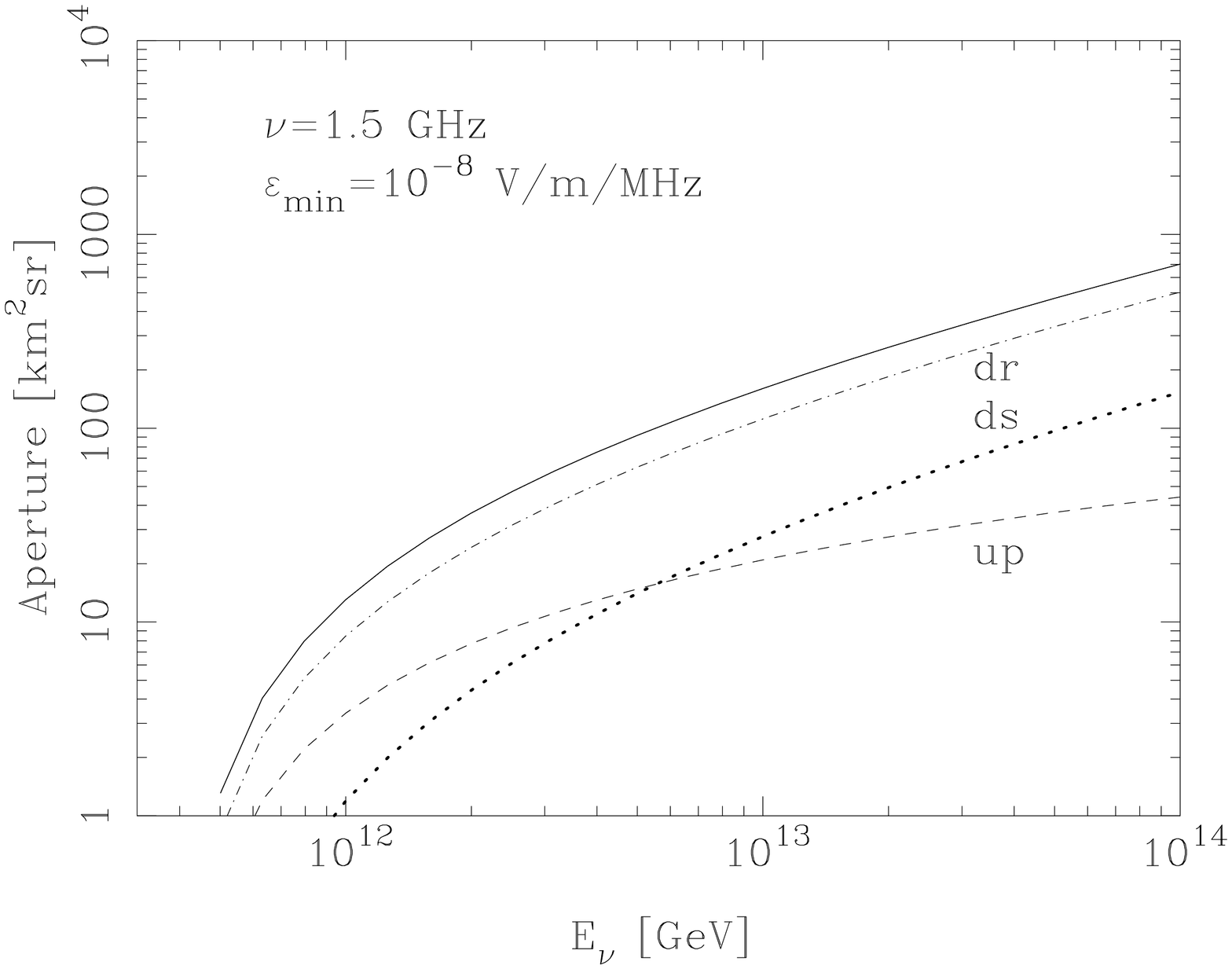}
\end{center}
\caption{The effective aperture as a function of neutrino
energy for $\varepsilon_{\rm min}= 10^{-8}$ V/m/MHz and for 
two frequency choices:  (a) $\nu=0.15$ GHz
and (b) $\nu=1.5$ GHz. The solid line shows the total of the downward rough (dr, dot-dashed), downward smooth (ds, dotted) and upward neutrino (up, dashed) contributions. }
\label{fig:effap8}
\end{figure}

\begin{figure}[t]
\begin{center}
\includegraphics[angle=270,width=0.45\textwidth]{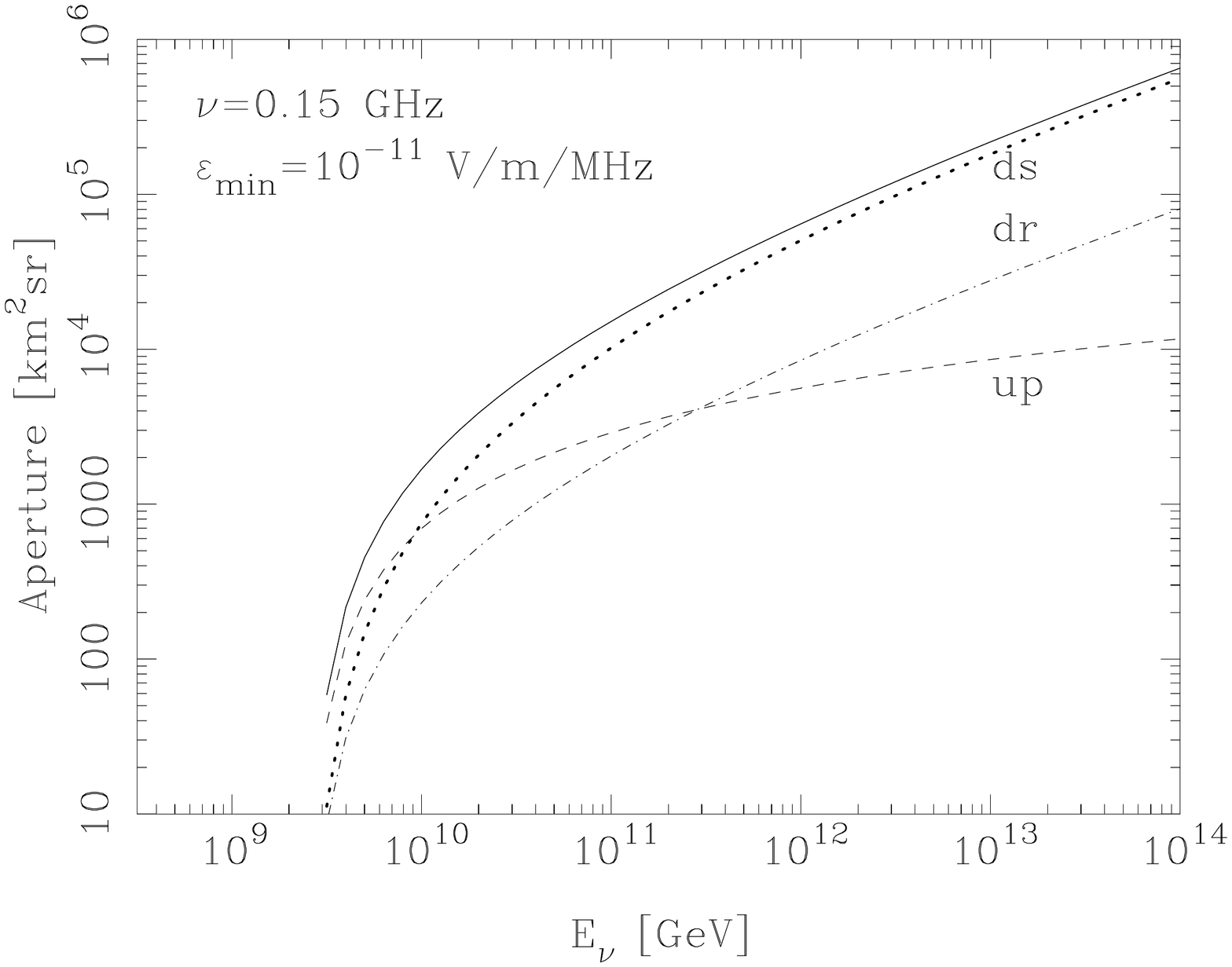}
\includegraphics[angle=270,width=0.45\textwidth]{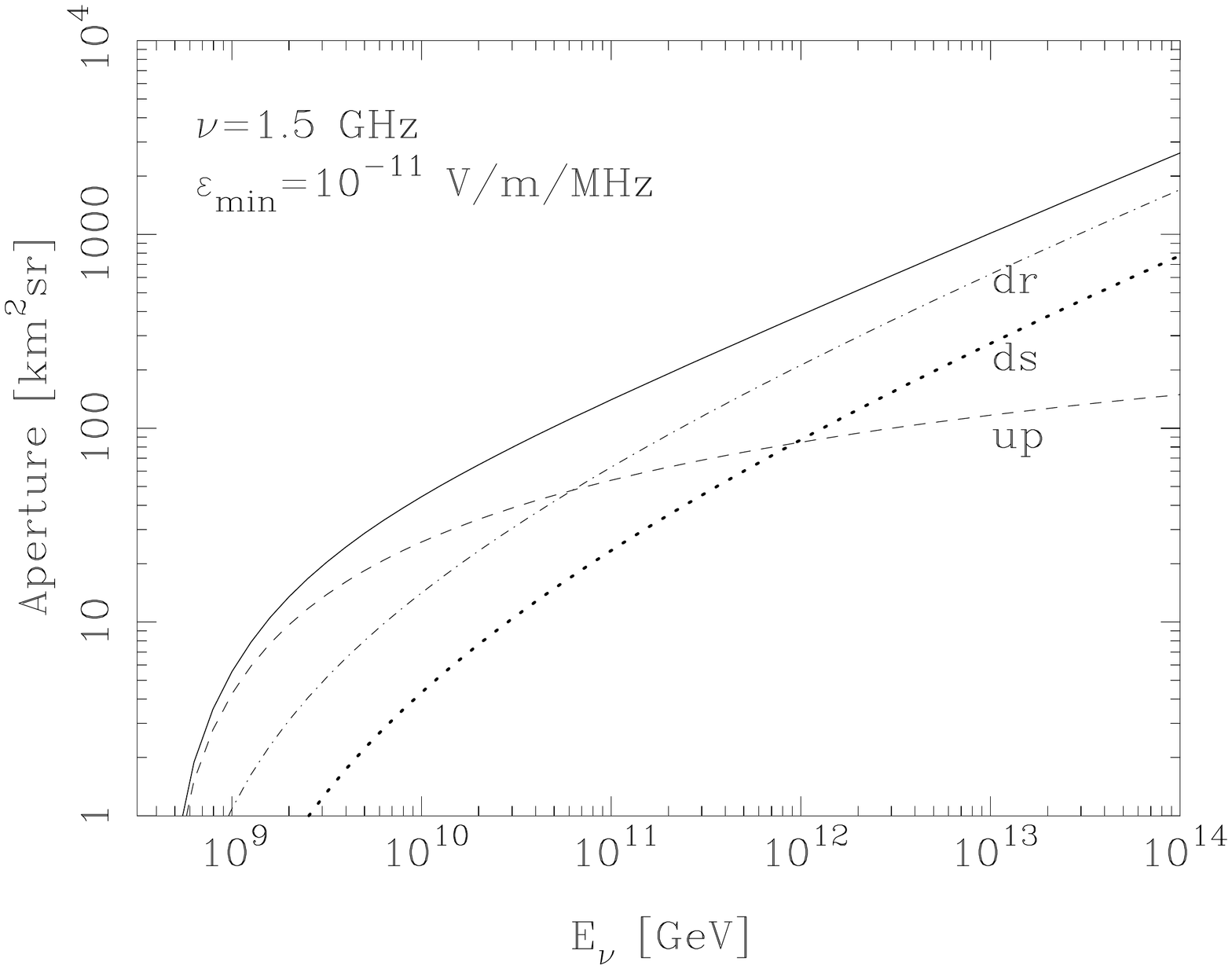}
\end{center}
\caption{The same as Fig. 5 but for 
$\varepsilon_{\rm min}= 10^{-11}$ V/m/MHz.} 
\label{fig:effap11}
\end{figure}

In Fig. \ref{fig:effap8}, we show the effective aperture for (a) $\nu=0.15$ GHz and (b) $\nu=1.5$ GHz,
for $\varepsilon_{\rm min} = 10^{-8}$ V/m/MHz.  For the lower frequency, the downward neutrinos in the smooth approximation dominate,
while for the higher frequency, roughness on the lunar surface transmits radio signals that would otherwise be lost.
At the lower frequency end, the angular spread Cherenkov cone is broader than at higher frequencies, allowing more of downward signal
to head towards Earth, even with an approximately smooth lunar surface. For the higher frequencies, the downward signals that emerge 
from the smooth surface approximation (ds) are smaller than the downward signals accounting for surface roughness. This comes from $f_0\Delta_0\sim 0.01-0.1$ which
is small compared to the solid angle characterizing the surface roughness, $\sigma_0\simeq 0.2$, as discussed by Gayley et al. \cite{gayley}.
In Fig. \ref{fig:effap11},
for 
$\varepsilon_{\rm min}= 10^{-11}$ V/m/MHz, 
we show the effective aperture for the same two radio frequencies. The lower minimum detectable
electric field allows a probe of a lower neutrino energy, where the flux of
neutrinos is predicted to be much larger.

\begin{figure}[h]
\begin{center}
\includegraphics[angle=270,width=0.45\textwidth]{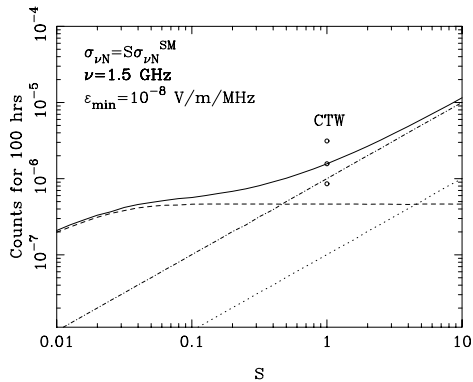}
\end{center}
\caption{The number of events for a 1.5 GHz signal with $\varepsilon_{\rm min}=10^{-8}$
V/m/MHz with the high cosmogenic flux from Fig. \ref{fig:flux}, with $E_\nu<10^{14}$ GeV, as a function of $S=\sigma_{\nu N}/
\sigma_{\nu N}^{SM}$. The up, down smooth and down rough contributions are as in Fig. \ref{fig:effap8} (b).
{ The dots labeled CTW show the uncertainty bands of Ref. \cite{ctw}.}}
\label{fig:rate8hi}
\end{figure}

\begin{figure}[h]
\begin{center}
\includegraphics[angle=270,width=0.45\textwidth]{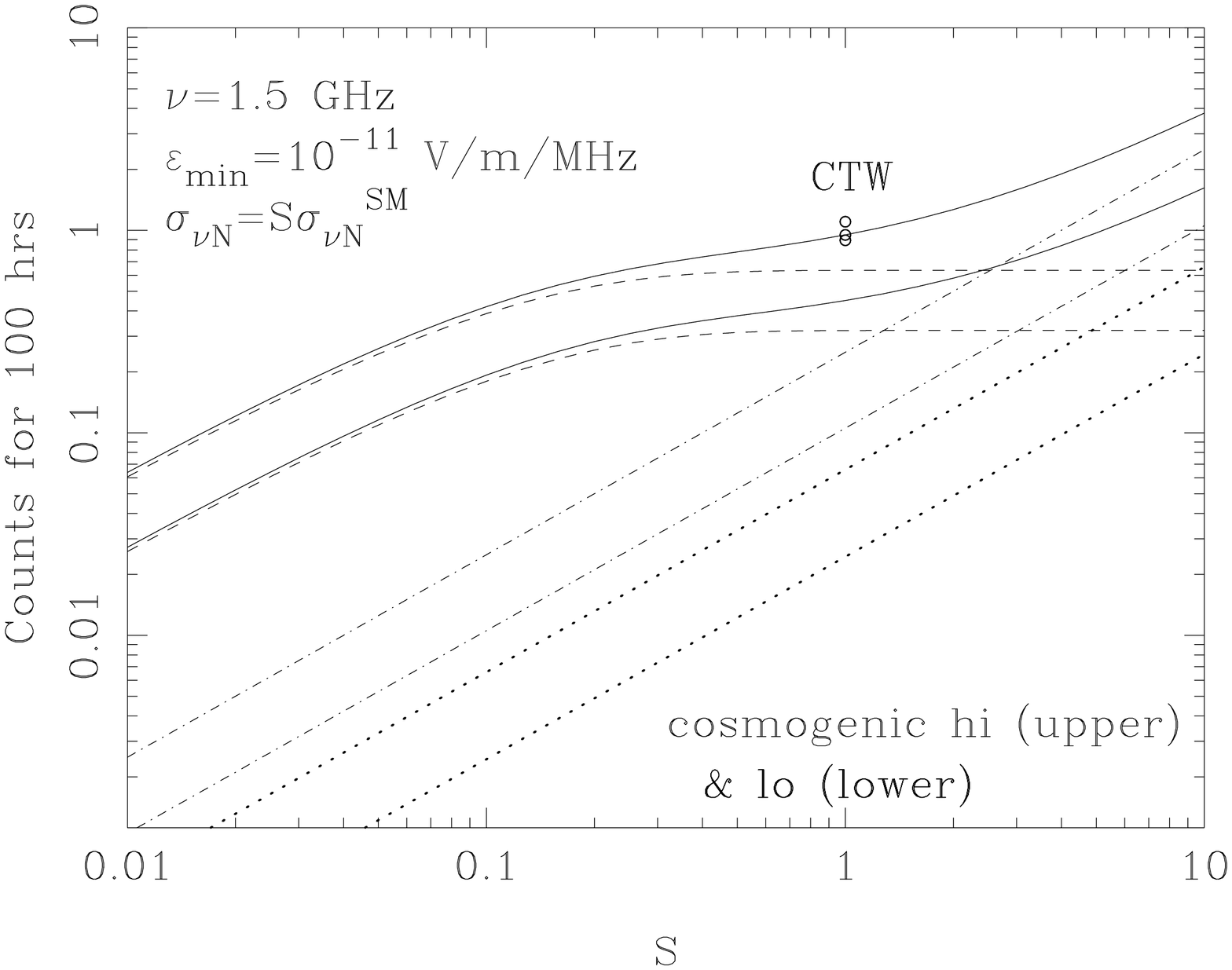}
\end{center}
\caption{The number of events for a 1.5 GHz signal as in Fig. \ref{fig:rate8hi},
with $\varepsilon_{\rm min}=10^{-11}$
V/m/MHz with the low and high cosmogenic flux from Fig. \ref{fig:flux}, with $E_\nu<10^{14}$ GeV, as a function of $S$.}
\label{fig:rate11}
\end{figure}

To show the dependence of the event rates on the neutrino nucleon cross section,
we take as an example
the radio frequency of $\nu=1.5$ GHz and 100 hrs of viewing time.
In Figs. \ref{fig:rate8hi} and \ref{fig:rate11}, we use a cross section
scaling factor of $S$ of the standard model (SM) neutrino nucleon cross section,
\begin{equation}
\sigma_{\nu N}= S \sigma_{\nu N}^{SM}
\ ,
\end{equation}  
to evaluate the
rates for two choices of $\varepsilon_{\rm min}$,  $\varepsilon_{\rm min}=10^{-8}$
V/m/MHz and $\varepsilon_{\rm min}=10^{-11}$
V/m/MHz. For the higher electric field threshold,
we have shown results only for the highest cosmogenic neutrino flux in Fig. 
\ref{fig:flux}.
For the lower threshold, we show results for both the upper and lower cosmogenic flux predictions.

Each of the separate contributions are shown: the dotted line show the ``down smooth''
contribution, the dot-dashed line shows the ``down rough'' contribution and
the dashed line shows the ``up'' contribution. As discussed, the ``up'' contribution
becomes independent of neutrino nucleon cross section when the cross section is 
large enough, while the ``down'' contributions scale linearly with the cross section.
For low cross sections, attenuation of the upward neutrino flux is less prominent,
so, for example, with $\varepsilon_{\rm min}=10^{-11}$
V/m/MHz and $S\sim 0.01$, the upward event rate scales with the neutrino nucleon cross section.

The total numbers of events in one hundred hours are shown with the solid
curves. The dots labeled CTW in each of the figures show a range of predictions
using the uncertainty bands of Connolly, Thorne and Waters (CTW) discussed
in Ref. \cite{ctw}. The uncertainty is largest at the highest energies, probed with the
lower electric field threshold. While the range of predictions spans a factor of about
five, the overall predicted rate is quite low, even for the higher cosmogenic
flux. The standard model
cross section (or the flux) must be enhanced by at least six orders
of magnitude to get one predicted event in 100 hours. If it is the 
cross section that is enhanced to this degree, the neutrino interaction
length becomes small compared to $L_\gamma$ and the approximations used
here for neutrinos do not apply. We discuss this possibility in the next section.

\begin{figure}[t]
\begin{center}
\includegraphics[angle=270,width=0.45\textwidth]{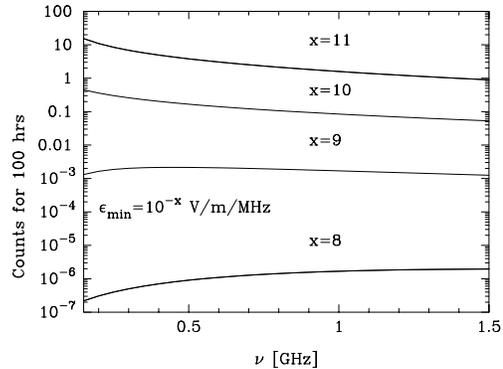}
\end{center}
\caption{The number of { neutrino}
events as a function of radio
frequency from {cosmogenic} neutrinos (high flux) { with standard model
interactions for} $\varepsilon_{\rm min}=10^{-11}-
10^{-8}$
V/m/MHz. }
\label{fig:freq}
\end{figure}

If an electric field detection threshold can be as low as $\varepsilon_{\rm min}=10^{-11}$
V/m/MHz, then on the order of one event is predicted { for $\nu=1.5$ GHz.} The standard model uncertainty
is much less in the energy regime probed by this electric field sensitivity.
Finally, we show the frequency dependence of eq. (\ref{eq:evtrate}) using
the standard model cross section for neutrinos and the high cosmogenic
neutrino flux. In Fig.
\ref{fig:freq}, we show the predicted number of events for 100 hours as
a function of detected radio frequency. For standard model neutrino
nucleon cross sections and the cosmogenic neutrino flux, the electric field
detection threshold at Earth must be on the order of $\varepsilon_{\rm min}=10^{-11}-
10^{-10}$ V/m/MHz for even one event in 100 hrs.

\section{Lunar Radio Cherenkov signals from cosmic ray protons}

In the previous section, we saw that for a range of energies, depending on detection parameters,
the downward neutrinos dominate the event rate. 
Contributions in the smooth case can dominate for some energies and for lower radio frequencies,
while the surface roughness is important for the higher radio frequencies. Given that downward production of
hadronic showers by neutrino interactions in the lunar regolith can produce observable radio
Cherenkov signals at Earth, one should also consider the corresponding signals from hadronic
showers induced by cosmic rays, the topic of this section. The Westebork group has already
used the absence of a cosmic ray induced radio Cherenkov signal to put a limit on the cosmic ray flux \cite{scholten}.

Our starting point is to treat the cosmic rays incident on the lunar regolith
with a flux shown in Fig. \ref{fig:flux}. The Pierre Auger Cosmic Ray Observatory analysis favors a composition of primarily
iron nuclei { at ultrahigh energies} \cite{iron}. 
Iron nuclei incident with the same cosmic ray energy as a single proton will
produce similar hadronic showers, so we use the cosmic ray flux of 
Ref. \cite{crflux} assuming that the incident particles are
protons carrying all the energy. 
For neutrino induced hadronic showers, we approximate the hadronic
shower energy to be $E_{shr}\simeq 0.2 E_\nu$. We will make the same approximation for cosmic
rays, $E_{shr}\simeq 0.2 E_{CR}$. 

The essential difference between incident cosmic rays and
neutrinos is the difference between strong interaction and weak interaction cross sections. For $E_\nu=
10^{12}$ GeV, the neutrino interaction length in the standard model is on the order of $10^7$ cmwe, as indicated
in Fig. 2, however, for cosmic rays, the interaction length is $L_{CR}\simeq 50$ cmwe. This new scale makes some of
the analytic approximations in Ref. \cite{gayley} { inapplicable} to cosmic rays.

The short interaction length of cosmic rays makes attenuation of the flux important for downward cosmic rays, and
it completely extinguishes the upward flux. We consider here the modifications to the effective aperture
for the downward flux contribution, including attenuation in the regolith. Details
of our evaluation appear in Appendix A.

Without attenuation, the 
target volume integral is governed by the maximum depth related to the photon attenuation length $L_\gamma$. 
In Ref. \cite{gayley}, the integral over the diameter of the moon $r$ is replaced by an integral over the 
perpendicular distance to the surface $h$. In the small angle approximation, the 
maximum depth is approximately
\begin{equation}
{h_{max}}\simeq L_\gamma\sin\theta_c f_0^2 \Biggl(1-\frac{\Delta^2}{f_0^2\Delta_0^2}\Biggr)\ ,
\end{equation}
where $\Delta$ is the polar angle from the Cherenkov peak.
(See eq. (13) in Ref. \cite{gayley}.) This factor of $L_\gamma$ arises because of the requirement that the radio
Cherenkov signal emerge from the regolith. 
Without flux attenuation, the integral over $h$ simply contributes a factor
of $h_{max}$. These are some of the factors that appear in eq. (\ref{eq:aperture})
for neutrinos.

With flux attenuation, the integral over $h$ in the evaluation of the effective aperture
results in a factor of $|\sin\alpha| L_{CR}$, where $\alpha$
is the angle of the incident cosmic ray with respect to the horizontal, as in Ref. \cite{gayley}. This modification
of eq. (29) in Ref. \cite{gayley}, and subsequent approximate integration, yields 
\begin{equation}
\label{eq:pcr}
P_{CR}(E)\simeq \frac{\sqrt{n_r^2-1}(f_0 \Delta_0)^3}{12}\Biggl(1+\frac{3}{4}\frac{\sigma_0^2}{f_0^2\Delta_0^2}
\Biggr)\ .
\end{equation}
The first term in parenthesis is the smooth contribution, and the second term includes the additional contribution from
surface roughness. 

Eq. (\ref{eq:pcr}) shows that the probability for a cosmic ray to produce a signal is independent of the cosmic ray
cross section, as long as cosmic ray flux attenuation is important on the scale of $L_{\gamma}$. This comes from two compensating factors.
One factor of $L_{CR}$ comes from the limit on the depth of targets from which there a signal, not because of radio wave attenuation, but 
because cosmic rays do not penetrate deeper in the regolith. The second factor is proportional to $\sigma_{CR}\sim 1/L_{CR}$ from
the probability that the cosmic ray interacts to produce a radio Cherenkov signal.

\begin{figure}[t]
\begin{center}
\includegraphics[angle=270,width=0.45\textwidth]{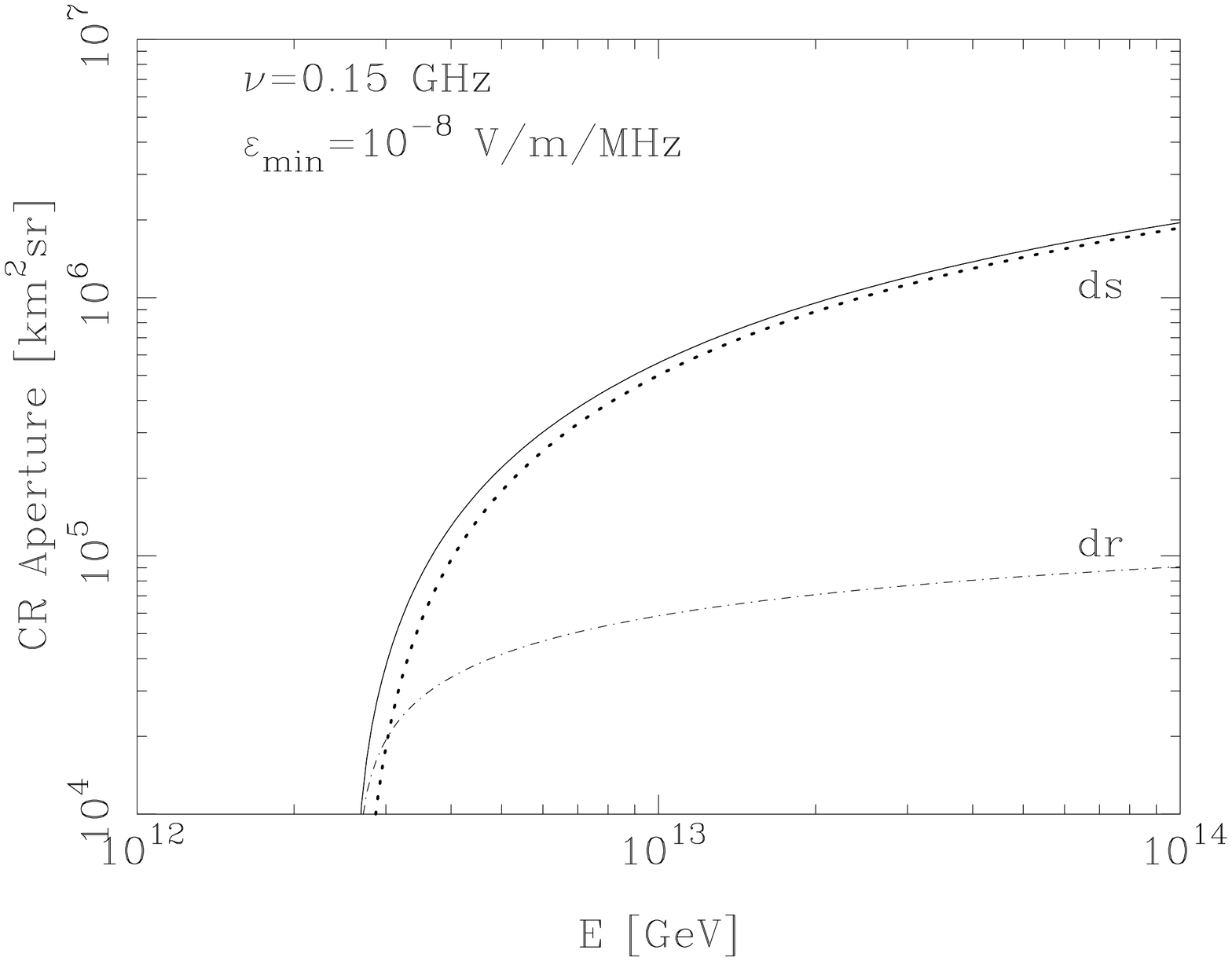}
\includegraphics[angle=270,width=0.45\textwidth]{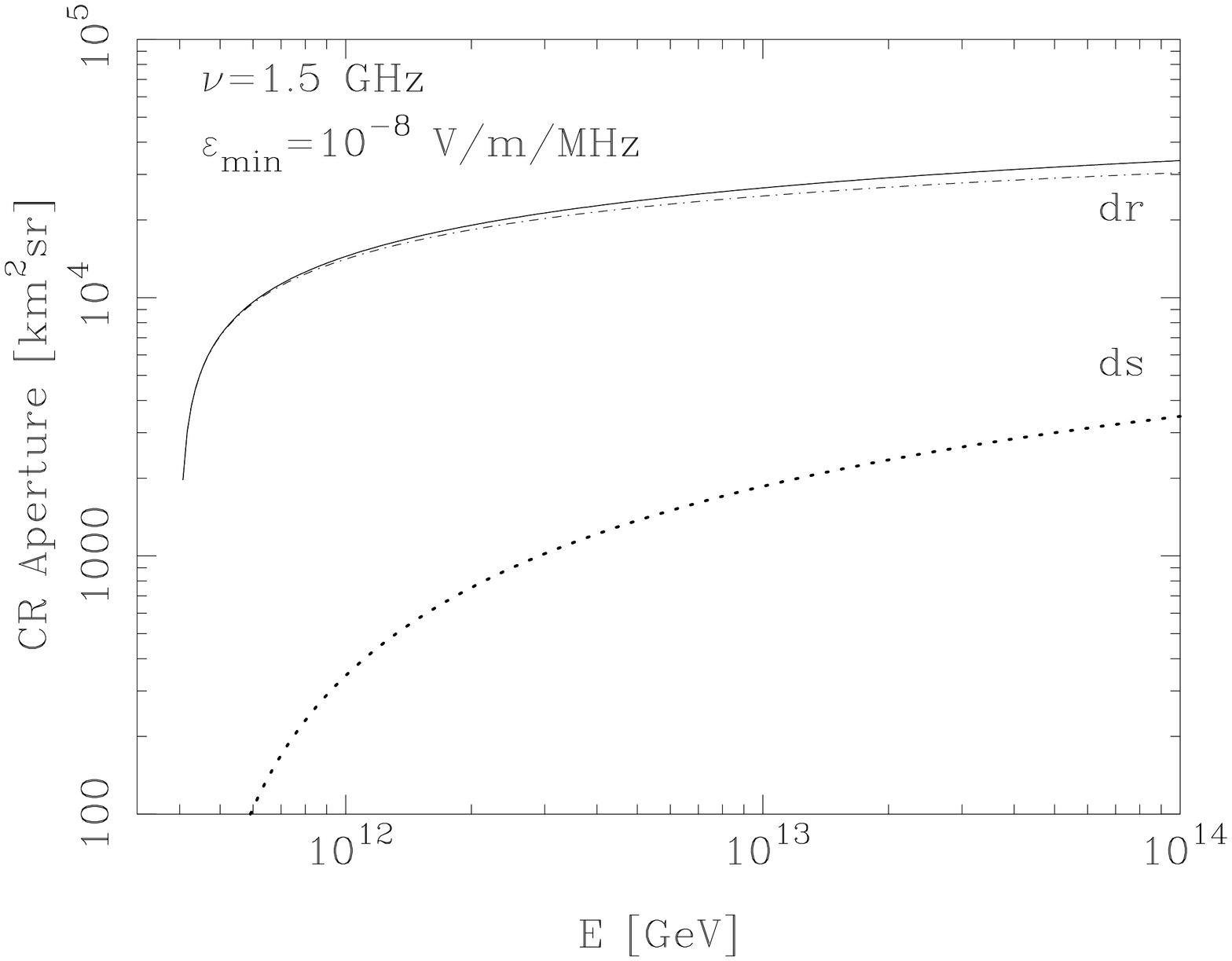}
\end{center}
\caption{The effective aperture as a function of cosmic ray 
energy for ($\varepsilon_{\rm min}= 10^{-8}$ V/m/MHz for (a) $\nu=0.15$ GHz
and (b)$\nu=1.5$ GHz. The solid line shows the total of the downward rough (dr, dot-dashed) and downward smooth (ds, dotted).The cosmic ray effective aperture is independent of the cosmic ray cross section.}
\label{fig:effapcr8}
\end{figure}

\begin{figure}[h]
\begin{center}
\includegraphics[angle=270,width=0.45\textwidth]{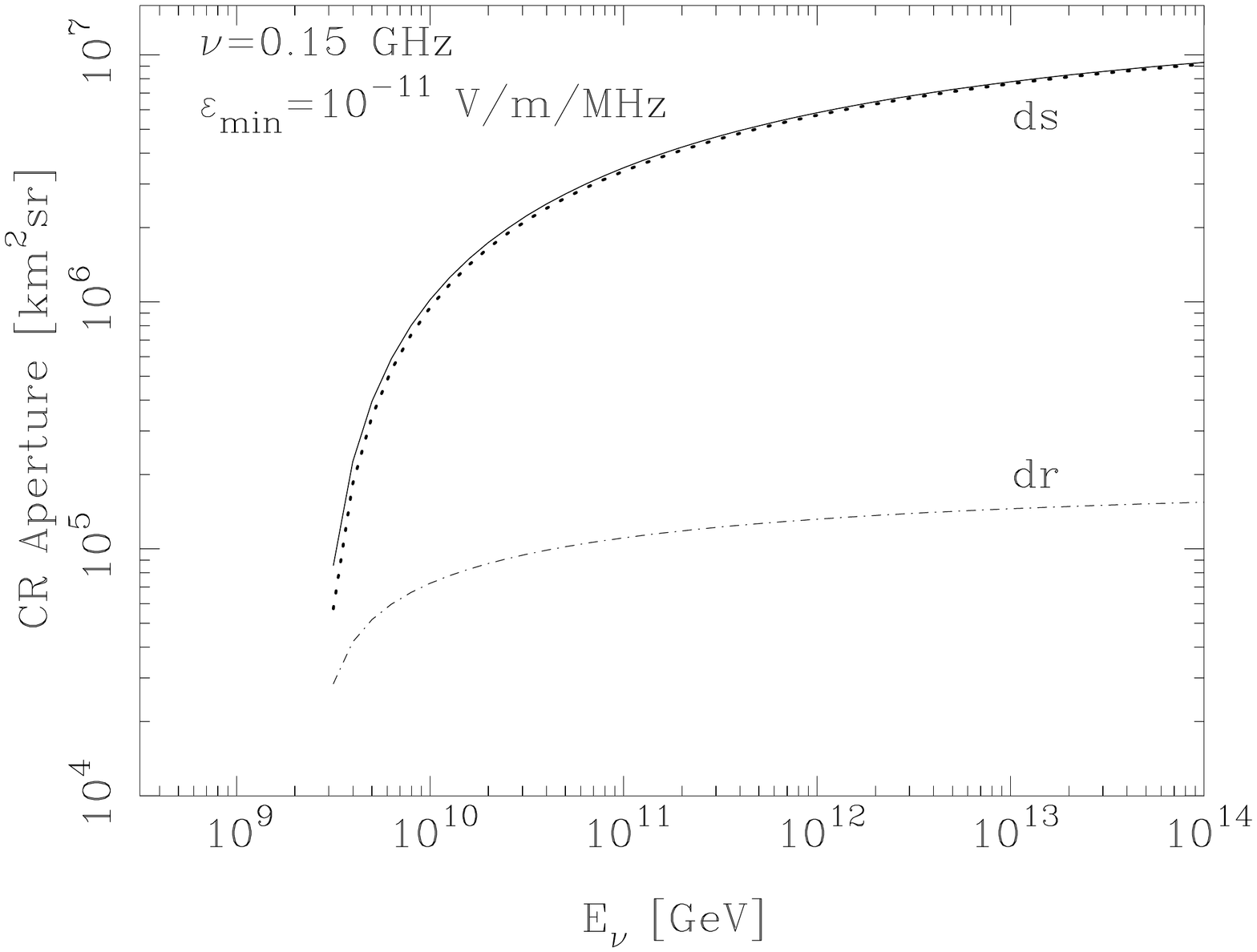}
\includegraphics[angle=270,width=0.45\textwidth]{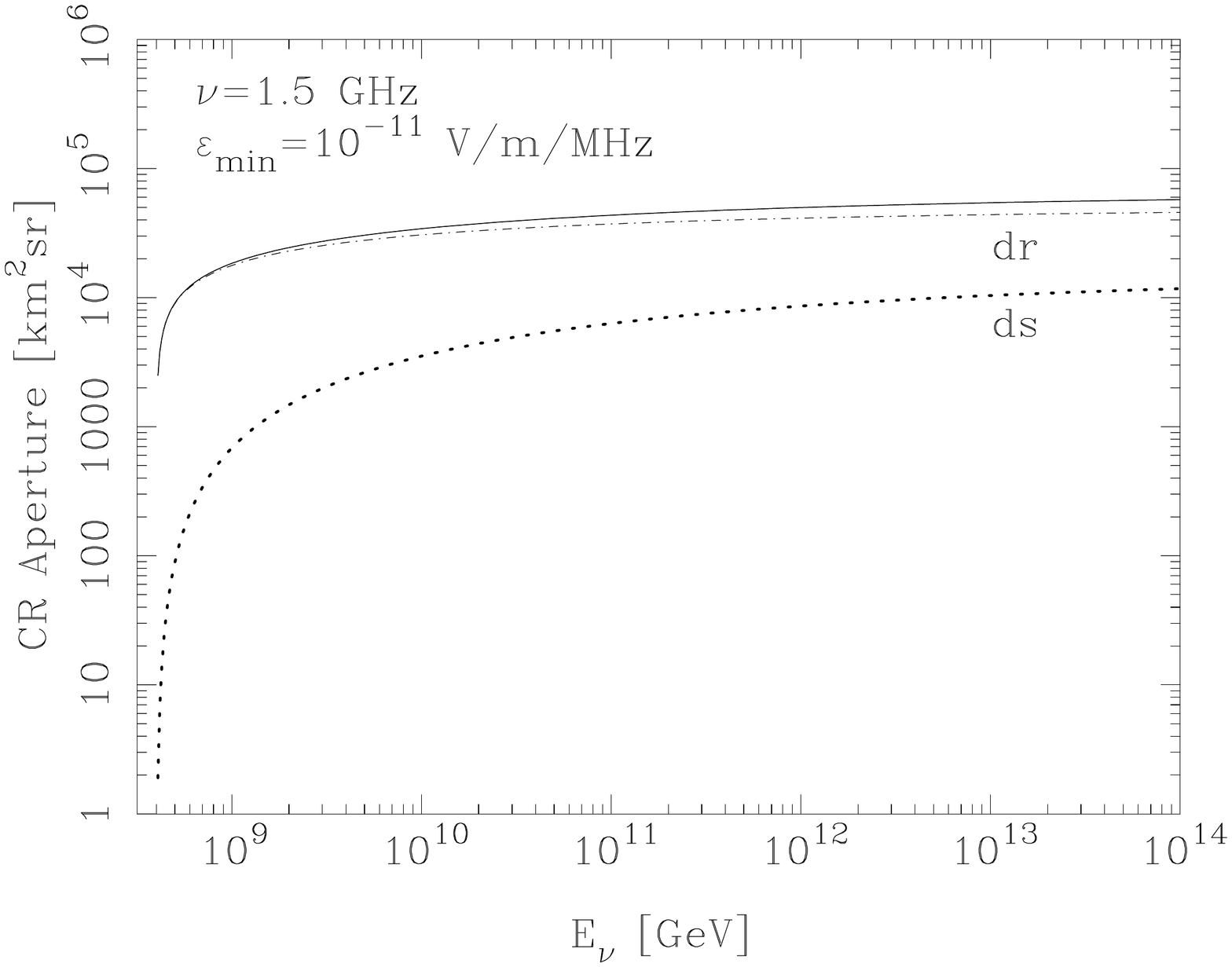}
\end{center}
\caption{The effective aperture as a function of neutrino
energy for ($\varepsilon_{\rm min}= 10^{-11}$ V/m/MHz for (a) $\nu=0.15$ GHz
and (b)$\nu=1.5$ GHz. The solid line shows the total of the downward rough (dr, dot-dashed) and downward smooth (ds, dotted).}
\label{fig:effapcr11}
\end{figure}

In Fig. \ref{fig:effapcr8}, we show the effective aperture for incident cosmic rays for $\nu=0.15$ and $1.5$ GHz and $\varepsilon_{\rm min}=10^{-8}$ V/m/MHz, and in Fig. \ref{fig:effapcr11}, we show the same for $\varepsilon_{\rm min}=10^{-11}$ V/m/MHz. We see a
similar dominance of the smooth contribution for the lower frequency, and rough contribution for the higher frequency, in the
cosmic ray induced radio Cherenkov signals.

\begin{figure}[h]
\begin{center}
\includegraphics[angle=270,width=0.45\textwidth]{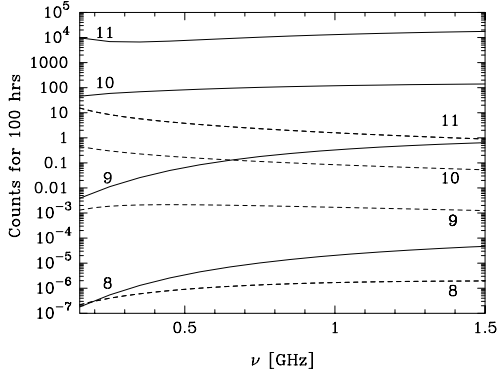}
\end{center}
\caption{The number of events as a function of radio
frequency from the high flux of cosmogenic neutrinos (dashed)
and from cosmic rays (solid), from top to bottom  with $\varepsilon_{\rm min}=10^{-11}-
10^{-8}$
V/m/MHz for standard model neutrino interactions. }
\label{fig:freqcr}
\end{figure}

Based on these results for the effective aperture, 
we evaluate the number of events as a function
of radio frequency induced by cosmic ray interactions. They are shown in Fig. \ref{fig:freqcr} with the solid lines. For reference, the neutrino induced
Cherenkov rates are shown with dashed lines. In all cases, the cosmic ray induced
rates are larger than the neutrino induced rates. This is true at low and high
radio frequencies, where either the smooth or rough downward contributions dominate. Our conclusion is that in the standard model of neutrino interactions
with nucleons, the cosmic ray induced Cherenkov signals will overwhelm the cosmogenic
neutrino induced signals, at least for the sample fluxes shown here.  The lower
electric field thresholds sample lower neutrino energies where there are larger
fluxes of neutrinos, but there are even larger fluxes of cosmic rays at those 
energies. The higher energy thresholds suffer from low rates due to the low flux.

Theoretical predictions for neutrino induced event rates are enhanced 
if either the neutrino flux is much larger than shown, for example, in Fig.
\ref{fig:flux}, or if the neutrino cross section is much larger than the
standard model cross section. It is this second possibility that we explore
in the next section.

\section{Radio Cherenkov signals and enhanced neutrino cross sections}

Enhanced neutrino-nucleon cross sections can arise in a variety of
extensions of the standard model. Here, we consider the case of
large extra dimensions.
In theoretical models with large extra dimensions and low scale 
gravity there is a possibility of creating a microscopic black 
hole in neutrino-nucleon interactions at very high energies \cite{add}.  
In these models, gravitational interactions 
are modified and the four dimensional Planck scale ($M_{Pl}$) is 
related to the fundamental Planck scale in $4+N_D$ dimension ($M_D$) by 
$$M_{\rm Pl}^2 = M_D^{N_D+2} V_{N_D}$$ 
where $V_{N_D}=(2\pi R)^{N_D}$ is the volume of the $N_D$-torus and $R$ is the size
of extra dimensions \cite{add,aadd}.  When $R$ is large (of the order of a millimeter), and 
the number of extra dimensions $N_D$ is larger than $2$, 
the fundamental Planck scale 
can be of the order of few TeV.  

In high energy collisions, 
when particles with energies above $M_D$ approach each 
other at the impact parameter which is
less than the Schwarzschild radius in $4+N_D$ dimensions, they can form the
$4+N_D$ dimensional black hole with mass $M_{BH}$ 
\cite{fengbh,feng,bhneutrinos,bhcolliders,yoshida,ctw}.
The radius $r_S$ is given by  \cite{mperry}
\be
r_S=\frac{1}{\sqrt{\pi}}\frac{1}{M_D}\Bigg[\frac{M_{BH}}{M_D}\Bigg(\frac{8
\Gamma(\frac{N_D+3}{2})}{N_D+2}\Bigg)\Bigg]^{\frac{1}{N_D+1}}\ .
\ee
The neutrino-nucleon cross section for black hole production is given by \cite{feng}
\begin{equation}
\sigma (\nu N \to {\rm BH}) = 
\sum_i \int_{\frac{(M_{BH}^{\rm min})^2}{s}}^1 dx \,\, \hat\sigma_i^{BH}(xs)\,\, f_i(x,Q^2),
\end{equation}
where 
$\hat\sigma_i^{BH}$ is the neutrino-parton cross section given by \cite{bhneutrinos}
\begin{equation}
\hat\sigma (\nu j\rightarrow BH) = \pi r_S^2(M_{BH}=\sqrt{\hat s}) 
\theta (\sqrt {\hat s} - M_{BH}^{\rm min})\ ,
\end{equation}
$s$ is the center of mass energy squared, $s=2m_NE_\nu$,
and $f_i(x,Q^2)$
is the parton distribution function for parton $i$.  
For semiclassical approximation to be valid, we need 
 $M_{\rm BH}^{\rm min}\gg M_D$.

Current limits on the $M_D$ and $N_D$ come from collider data 
\cite{bhcolliders,cms} as 
well as from the astrophysical observations.  Strongest limits for 
$N_D<4$ come from supernova 
cooling and neutron star heating, $M_D>4$ TeV for $N_D=4$, 
$M_D>0.8$ TeV for $N_D=5$ and for lower values of $N_D$, $M_D$ 
is constrained to be much larger than few hundred TeV.  
Non-observation of the black hole production in 
cosmic neutrinos provide stringent limit for 
$N_D>5$, 
$M_D>1$ TeV \cite{feng}.  

We consider the parameter space ($M_D$, $N_D$), for 
fixed $M_{BH}^{min}$ using the cross sections from
Ref. \cite{ctw} where 
$M_{BH}^{min} = M_D$, to illustrate the consequences of
enhanced neutrino nucleon cross sections 
given the highest cosmogenic neutrino flux
shown in Fig. 1.  

\begin{figure}[t]
\begin{center}
\includegraphics[width=0.45\textwidth]{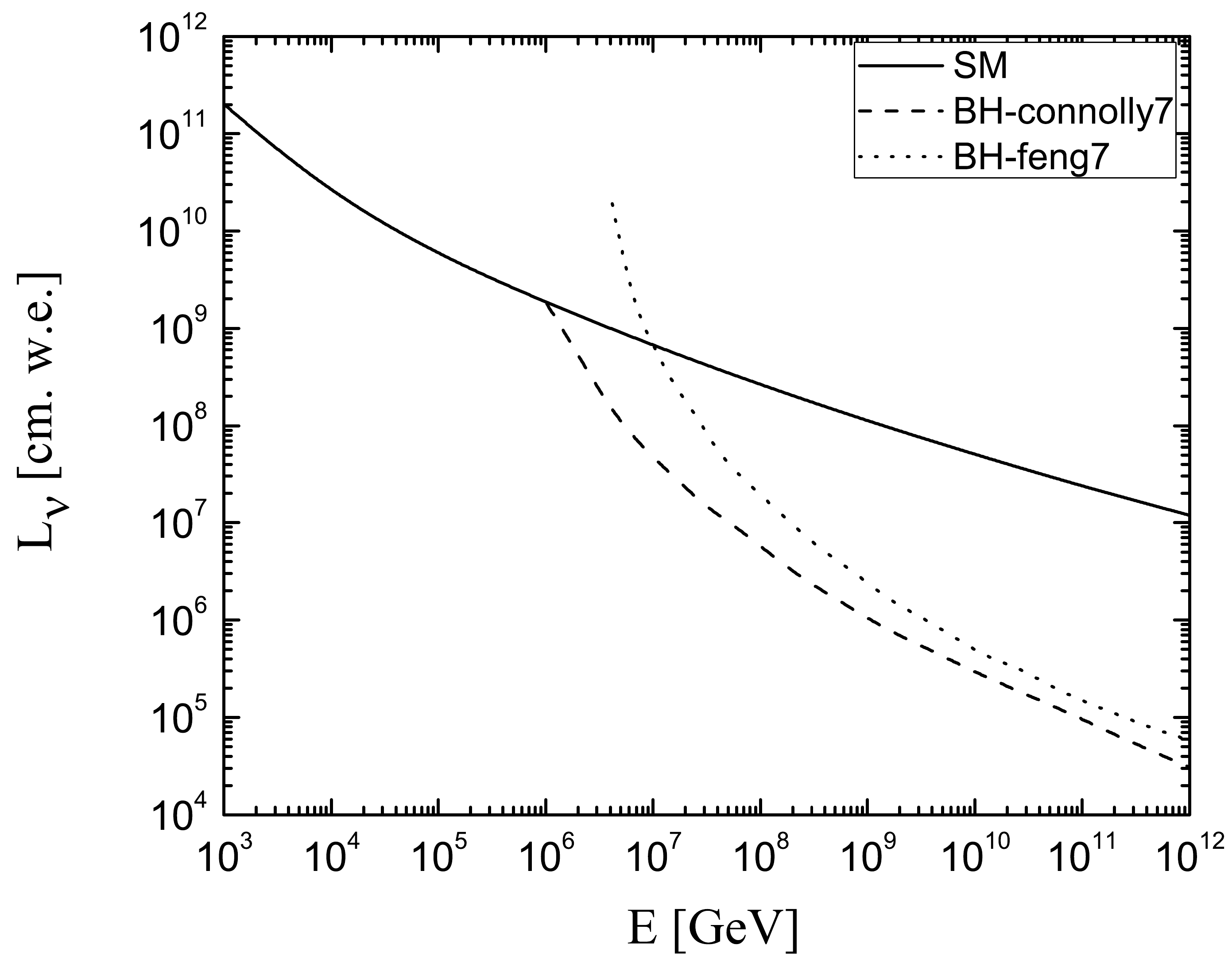}
\end{center}
\caption{The neutrino interaction length, with contributions from non-standard model mini-black hole production from Connolly, Thorne and Waters \cite{ctw} and
Feng \cite{fengbh} for $N_D=7$, $M_D=1$ TeV.}
\label{fig:intminibh}
\end{figure}

The neutrino-nucleon cross section for black hole production 
exceeds standard model cross section for neutrino energies 
above ~$10^6$ GeV and is about two orders of magnitude 
larger than the standard model cross section at $E_\nu \sim 10^{11}$ 
GeV.  In Fig. \ref{fig:intminibh} we show the effect  of the mini-black 
hole contribution to the neutrino-nucleon cross section on 
 neutrino interaction
length. The dashed line shows the interaction length for 
$N_D=7$ and $M_D=1$ TeV
as calculated in Ref \cite{ctw}.  The dotted line shows the interaction
length for the same ($N_D$, $M_D$) with
the cross section taken from Ref. \cite{fengbh}. We note that the neutrino interaction 
length due to black hole production had strong energy dependence and at 
neutrino energies above $10^9$ GeV, it is order of magnitude smaller 
than in the case of the standard model neutrino interactions.  
We use the  
neutrino cross sections due to black hole production 
from Ref. \cite{ctw} 
as 
sample cross sections in what follows below.

The fact that the neutrino interaction length decreases to less 
than $10^4 $ cmwe
above $E_\nu=10^{12}$ GeV means that neither the cosmic ray approximation of strong
attenuation
nor the standard model neutrino evaluation of the effective downward aperture applies.
The details of the evaluation of the effective aperture with the
mini-black hole enhanced neutrino cross section contribution to 
the probability function $P_{BH\nu}(E)$ is shown in
the Appendix.   

\begin{figure}[t]
\begin{center}
\includegraphics[angle=270,width=0.45\textwidth]{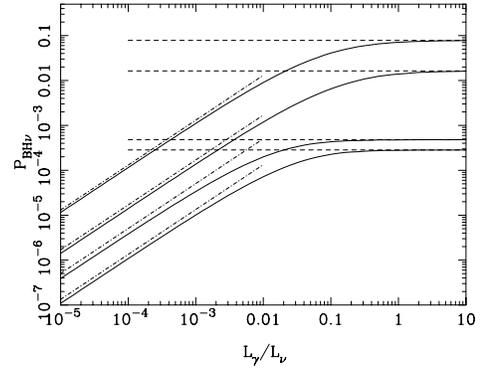}
\end{center}
\caption{The probability function which enters into the effective aperture, for neutrinos in which there is an enhancement of the neutrino cross section. The solid line shows the numerical evaluation, assuming $E=10^{14}$ GeV, but varying
the neutrino cross section relative to a 
fixed $L_\gamma$. The dashed horizontal 
line is the cosmic ray approximation of 
Eq. (\ref{eq:pcr}) and the dot-dashed line is the downward
neutrino approximation of Eq. (\ref{eq:aperture}). In order
from top to bottom, the curves show ($\varepsilon_{\rm min},\nu$) for
$(10^{-11},0.15), (10^{-8},0.15), (10^{-11},1.5)$ and $(10^{-8},1.5)$
in units of (V/m/MHz, GHz).
}
\label{fig:probnp}
\end{figure}

To illustrate dependence of $P_{BH\nu}(E)$ on 
the neutrino interaction length, we evaluate 
$P_{BH\nu}(E)$ as a function of the ratio of the $L_\gamma/L_\nu$. 
We show 
 in Fig. \ref{fig:probnp} { with the solid line} the
numerical evaluation of 
$P_{BH\nu}(E)$ { from the formulas in GMJ with the modification of
neutrino attenuation even over the distance scale of $L_\gamma$} . The dashed horizontal 
lines show the what could be called the
``cosmic ray limit, '' when the non-standard model
interactions make the neutrino interaction length
small compared to $ L_\gamma$.
The dot-dashed lines show
the ``standard model neutrino limit'' when $L_\nu\gg L_\gamma$ (the analytic
expression of GMJ). In order
from top to bottom, the curves show 
($\varepsilon_{\rm min},\nu$) for
$(10^{-11},0.15), (10^{-8},0.15), (10^{-11},1.5)$ and $(10^{-8},1.5)$
in units of (V/m/MHz, GHz).  
We note that the probability function for { enhanced
neutrino cross sections} approaches the cosmic ray approximation for 
 $L_\gamma/L_\nu \geq 0.1$ ($1.0$)  when 
$\varepsilon_{\rm min}=10^{-8}$ V/m/MHz 
($\varepsilon_{\rm min}=10^{-11}$ V/m/MHz)
and $\nu =1.5$ GHz 
($\nu =0.15$ GHz). 
For lower values of 
 $L_\gamma/ L_\nu$, the probability follows dependence 
on $L_\gamma/ L_\nu $ similar to the 
 the downward neutrino approximation 
of Eq. (12).  
As noted in Ref.
\cite{gayley}, the approximate analytic solutions reproduced in 
our Eq. (\ref{eq:aperture}) are good to better than 25\% compared to a numerical
solution of the multidimensional integrals in their analytic expression.
In evaluating neutrino induced Cherenkov signals with mini-black 
hole enhanced cross sections, we will use the numerical result 
(the solid line in Fig. \ref{fig:probnp}) to 
ensure that we are properly accounting for the transition 
from weak to strong neutrino cross sections.

\begin{figure}[t]
\begin{center}
\includegraphics[width=0.45\textwidth]{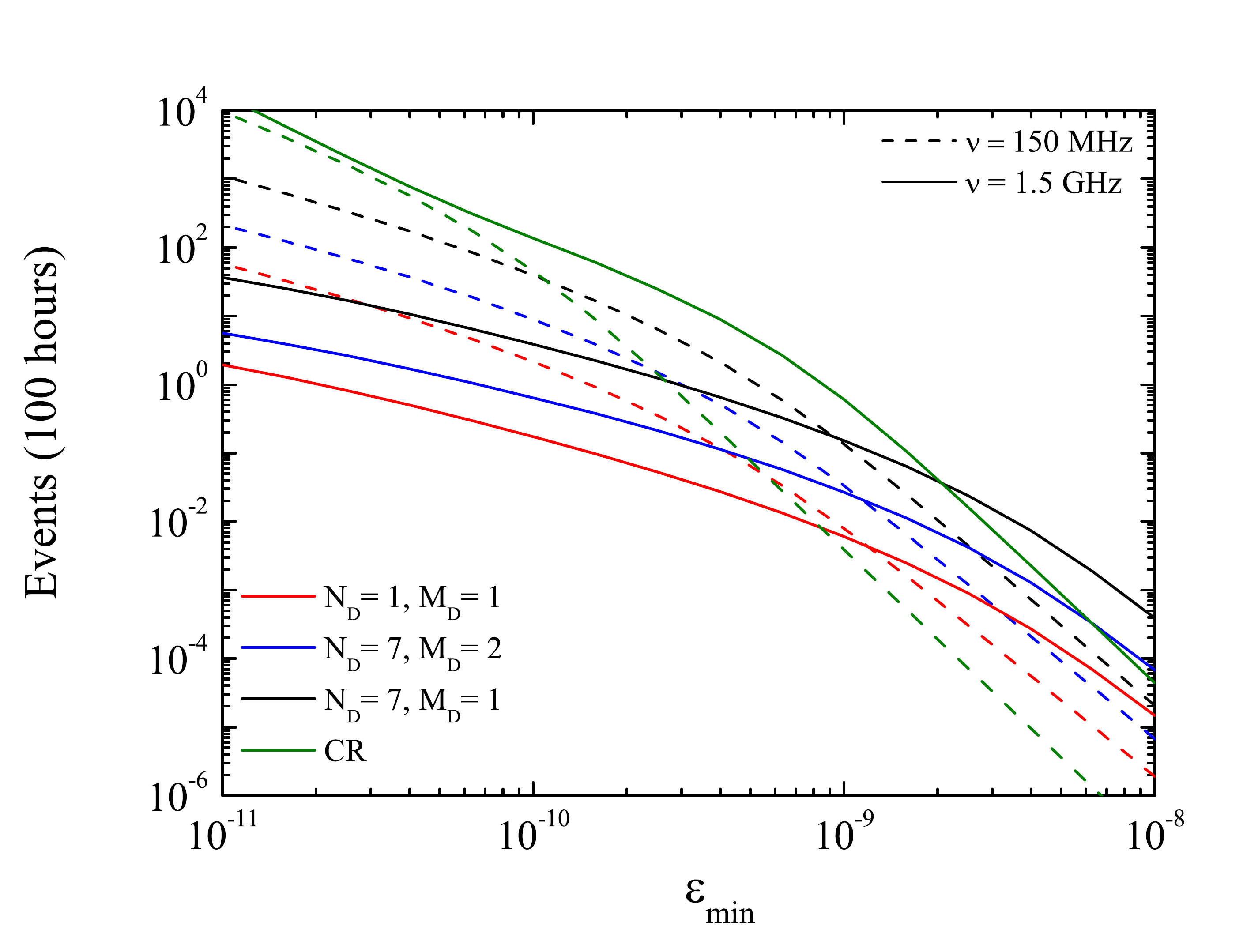}
\end{center}
\caption{Events in 100 hrs for $\nu=150$ MHz (dashed)
and 1.5 GHz (solid) as a function 
of $\varepsilon_{\rm min}$ (in V/m/MHz) using enhanced neutrino cross sections from 
 Connolly et al. \cite{ctw}, for three choices of black hole 
 parameters, $N_D$ and $M_D$.  The high cosmogenic flux from 
Fig. 1 is used.
The parameters ($N_D$, $M_D$) are
(1, 1 TeV), (7, 2 TeV) and 
(7, 1 TeV) from lowest to highest event rates respectively. The top solid and dashed curve at
$\varepsilon_{\rm min}=10^{-11}$ V/m/MHz are the cosmic ray rates.}
\label{fig:bhemin}
\end{figure}

In Fig. \ref{fig:bhemin} we show number of cosmogenic neutrino
events in 100 hrs for 
two frequency choices, $\nu=0.15$ GHz (dashed) and $\nu=1.5$ GHz (solid), 
as a function of the minimum electric
field $\varepsilon_{\rm min}$.   The parameters ($N_D$, $M_D$) are
(1, 1 TeV), (7, 2 TeV) and (7, 1 TeV)  from lowest to highest.
The highest pair of solid and dashed lines for
low $\varepsilon_{\rm min}$ is for the cosmic
ray induced events.
We find that for $\nu=1.5$ GHz (solid lines), 
the neutrino
induced BH contributions are larger
than the cosmic ray contributions for 
$\varepsilon_{\rm min}> 2\times 10^{-9}$ 
V/m/MHz 
($\varepsilon_{\rm min}> 6\times 10^{-9}$ 
V/m/MHz) 
for $N_D=7$ 
 and 
and $M_D=1$ TeV ($M_D=2$ TeV) black hole parameters.  
This comes from the highest energies considered, where the neutrino nucleon 
cross section is much bigger than the standard model, so the standard model neutrino 
interactions are negligible.  

For $\nu=0.15$ GHz, shown with the solid 
lines in Fig. \ref{fig:bhemin}, the neutrino 
induced BH contributions{ are larger than the cosmic ray 
induced Cherenkov signal for 
$\varepsilon_{\rm min}> 10^{-10}$ 
V/m/MHz, lower than the $\varepsilon_{\rm min}$ for $\nu=1.5$ GHz
required for a signal from neutrinos with}
$N_D=7$  
and $M_D=1$ TeV black hole parameters.  

From Fig. \ref{fig:bhemin}, we note that different BH parameters 
 give similar dependence of the events on minimum electric field, with only 
difference being in the overall number of events.  
With 100 hrs, the BH enhanced neutrino signals dominate the cosmic 
ray background 
in the range of 
40 events  at the crossover for
$\nu=0.15$ GHz, while for $\nu=1.5$ GHz, the crossover occurs at a 
fraction of an event level for the largest BH cross section considered here.

\begin{figure}[t]
\begin{center}
\includegraphics[width=0.45\textwidth]{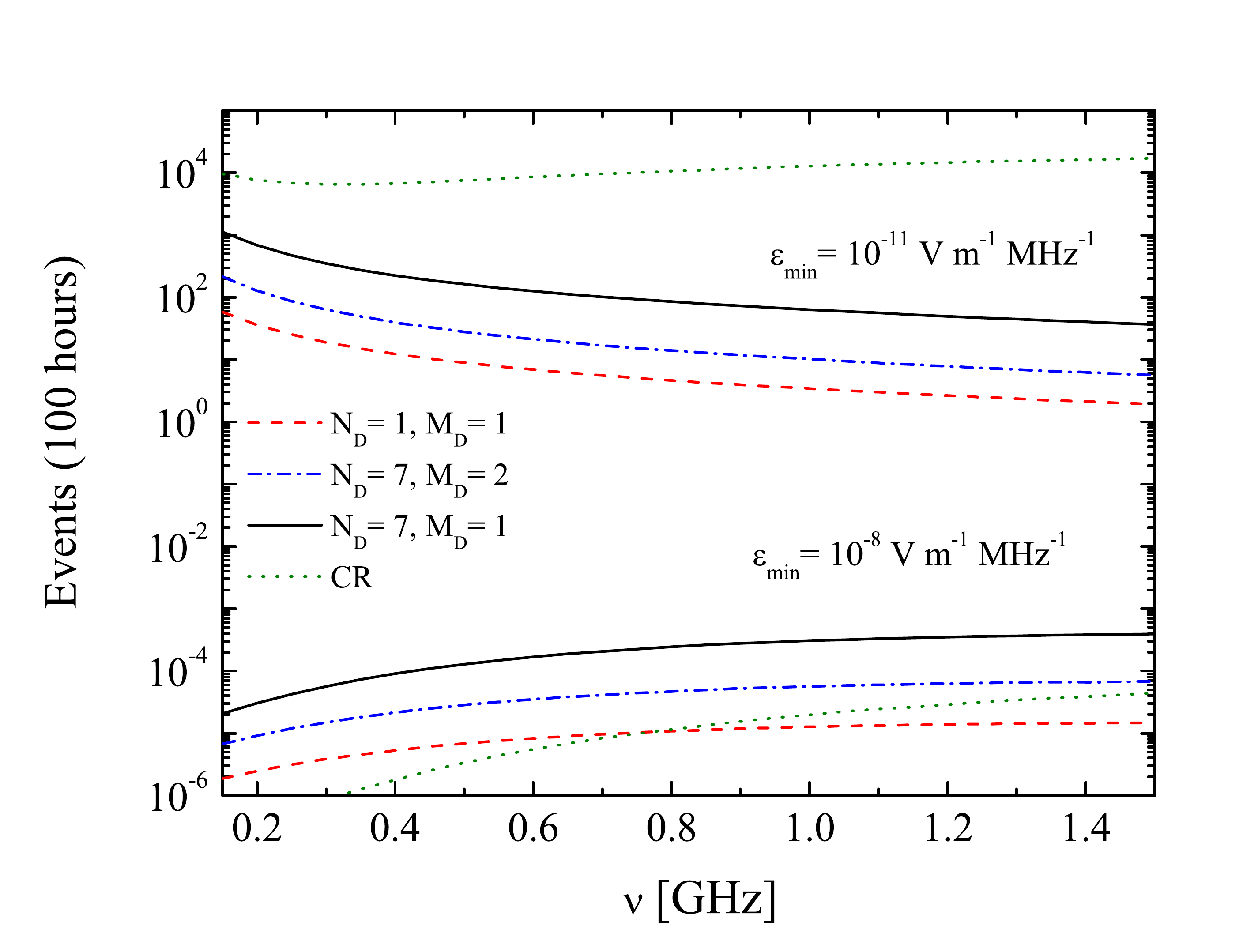}
\end{center}
\caption{Events in 100 hrs as a function of radio frequency for $\varepsilon_{\rm min}
= 10^{-11}$ V/m/MHz (upper curves) and $\varepsilon_{\rm min}
= 10^{-8}$ V/m/MHz (lower curves) for three choices of mini-blackhole parameters with the
high cosmogenic neutrino flux. The
cosmic ray induced event rate is shown with the dotted lines.}
\label{fig:bhfreq}
\end{figure}

In Fig. \ref{fig:bhfreq} we show the frequency dependence of the event rate
for the lower and higher minimum electric field for three values of ($N_D, M_D$).  There is 
roughly an order of magnitude change in the neutrino induced
event rates as a function of radio frequency for $\nu=0.15-1.5$ GHz.  
Again we see in the figure 
that while event rates are higher for lower minimum electric fields, 
the cosmic ray background is much larger than the signal and thus 
dramatically 
lowering minimum electric fields does not favor the
observation of cosmogenic neutrino induced Cherenkov signals. 

Our conclusion is that for cosmogenic neutrino fluxes and
neutrino cross sections enhanced by BH production with
$N_D=7$ and $M_D=1$ TeV, 
the neutrino signal equals
the cosmic ray background for $\varepsilon_{\rm min} \sim 10^{-10}$ V/m/MHz 
when $\nu=0.15$ GHz, and the number of events is about 40. Thus, 
improvements in detector sensistivies 
would be required to observe the neutrino black hole enhanced signal. 
For a current detector
capability of 
$\varepsilon_{\rm min}\sim 10^{-8}$ V/m/MHz, the 
enhanced
neutrino cross section dominates the 
radio Cherenkov signal from the Moon relative to the 
cosmic ray 
background, however, the number of events in 100 hours is 
too small, of the order of $4 \times 10^{-4}$.  

In addition to the cosmogenic neutrino flux that we have considered, 
 there is a possibility that 
astrophysical sources produce larger neutrino fluxes 
via 
Fermi shock acceleration of the charged particles such as protons 
which collide with other protons or photons in a disk or a jet 
producing mesons which decay into neutrinos.  
Neutrino fluxes due to shock acceleration follow a power-law, i.e.,
$\Phi_\nu \sim E^{-2}$.  
We consider a neutrino flux which is currently below the 
Anita limit \cite{Anita2}, i.e. ,
\begin{equation}
E^2\Phi_\nu = 5 \times 10^{-7} \ {\rm GeV/cm^2/s/sr}\ .
\label{eq:anitalimit}
\end{equation}

In Fig. \ref{fig:bhemin5em7} 
 we show the event rates 
for the neutrino flux in eq. (\ref{eq:anitalimit}) and 
for neutrino enhanced cross sections 
due to black hole production obtained with different black hole parameters 
($N_D$, $M_D$) from Connolly et al. \cite{ctw}, 
 as a function of the minimum electric 
field.  We find that even for 
$ \varepsilon_{\rm min}\sim 10^{-8}$ V/m/MHz, which is { a current detector capability}, for any frequency above $150$ MHz 
the signal is above the cosmic background by several orders of 
magnitude.  
Total number of events varies between $1$ (for 
$N_D=1$ and 
$M_D=1$ TeV) and 
$20$ (for 
$N_D=7$ and 
$M_D=1$ TeV) 
for $\nu = 1.5$ GHz and between $40$ and $200$ for 
$\nu =150$ MHz, depending on BH parameters.  Lowering 
$\varepsilon_{\rm min}$ results in only a slight increase in
the event rates.  For 
$\varepsilon_{\rm min} > 4 \times 10^{-10}$ 
the neutrino signal is above the cosmic 
ray background for any choice of BH parameters that we considered.

\section{Discussion}

We have used the formalism of GMJ in Ref. \cite{gayley} to evaluate the effective 
aperture for lunar observations from Earth of radio Cherenkov signals { produced by
neutrino interactions}. The GMJ analytic effective aperture agrees qualitatively with the 
Monte Carlo results of James and Protheroe \cite{jp}, as discussed in Ref. \cite{gayley}.
In the analytic approach, we have made a new evaluation of the lunar effective aperture for cosmic rays.
This relies on
the result of Ref. \cite{scholten}
that cosmic ray interactions in the lunar regolith produce radio Cherenkov signals that are
indistinguishable from neutrino induced signals.

As Fig. 12 shows, the cosmic ray induced event rate always
dominates the standard model neutrino event rate for the cosmogenic models of Ref. \cite{olinto}
for $\varepsilon=10^{-8}-10^{-11}$ V/m/MHz. Current capabilities
for the electric field { threshold are on the order of} 
$\varepsilon_{\rm min}\sim 10^{-8}$ V/m/MHz. Decreasing $\varepsilon_{\rm min}$ will not improve the
capability of this lunar technique to detect the cosmogenic neutrino flux if the standard model interactions,
 e.g., with cross sections on the order of
those reported in Refs. \cite{gqrs} and \cite{jeong1},
are the only interactions responsible for the neutrino cross section. While lowering the minimum electric
field increases the neutrino induced signal, it also increases the cosmic ray induced signal.
The effective aperture for cosmic rays is independent of the cosmic ray cross section.
Thus, the current discussion about the cosmic ray composition at the highest energies \cite{iron}
will not impact our results as long as the cosmic rays interact strongly. 

The radio Cherenkov technique can be exploited for lunar observations if the neutrino nucleon cross section
is increased. We showed that for the choice of $M_D=1$ TeV and $N_D=7$ for BH production,
an observable neutrino induced radio Cherenkov signal from the Moon is induced and it is larger than 
the cosmic ray induced background. For $\nu=150$ MHz, this would require an instrumental
improvement to reach $\varepsilon_{\rm min}\sim 10^{-10}$ V/m/MHz for the high cosmogenic flux
of Fig. 1. We note that eventually enhancements of the neutrino cross section will push the
rate evaluation into the ``cosmic ray regime,'' where the rate is independent of the cross section.

\begin{figure}[t]
\begin{center}
\includegraphics[width=0.45\textwidth]{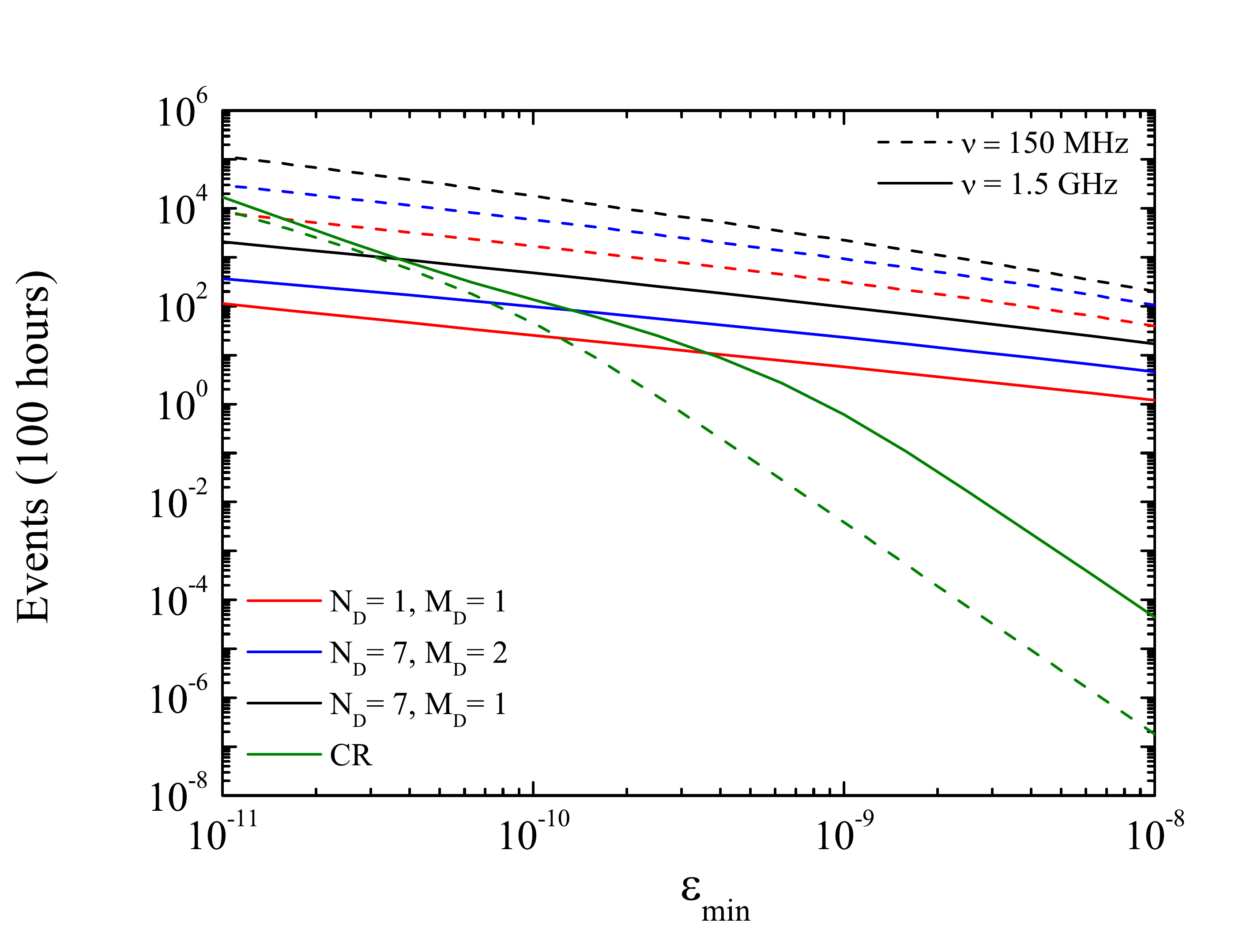}
\end{center}
\caption{Events in 100 hrs as a 
function of $\varepsilon_{\rm min}$ for the enhanced neutrino cross section 
due to black hole production, for different values of 
($N_D$, $M_D$) as in Fig. \ref{fig:bhemin}
and $E^2\Phi_\nu=5\times 10^{-7}$ GeV/cm$^2$/s/sr. 
The lowest solid and dashed curves at
$\varepsilon_{\rm min}=10^{-8}$ V/m/MHz are the cosmic ray rates.}
\label{fig:bhemin5em7}
\end{figure}

In Fig. \ref{fig:bhemin5em7}, we showed that if both the neutrino cross section and the neutrino flux are enhanced,
lunar radio Cherenkov techniques can make observations rather than set limits,
even for $\varepsilon_{\rm min}\sim 10^{-8}$ V/m/MHz. { Fig. \ref{fig:bhemin5em7} shows
the observational capability as a function of $\varepsilon_{\rm min}$ for two frequencies 
at the current limit on a neutrino flux of $E^2\Phi_\nu=5\times 10^{-7}$ GeV/cm$^2$/s/sr
for 
a specific mini-blackhole model of neutrino cross section. In this model, the cross section enhancement
relative to the standard model is 
$S=\sigma_{\nu N}/\sigma_{\nu N}^{SM}
\sim 500$ for $E_\nu=10^{12}$ GeV.}

\begin{figure}[h]
\begin{center}
\includegraphics[width=0.45\textwidth]{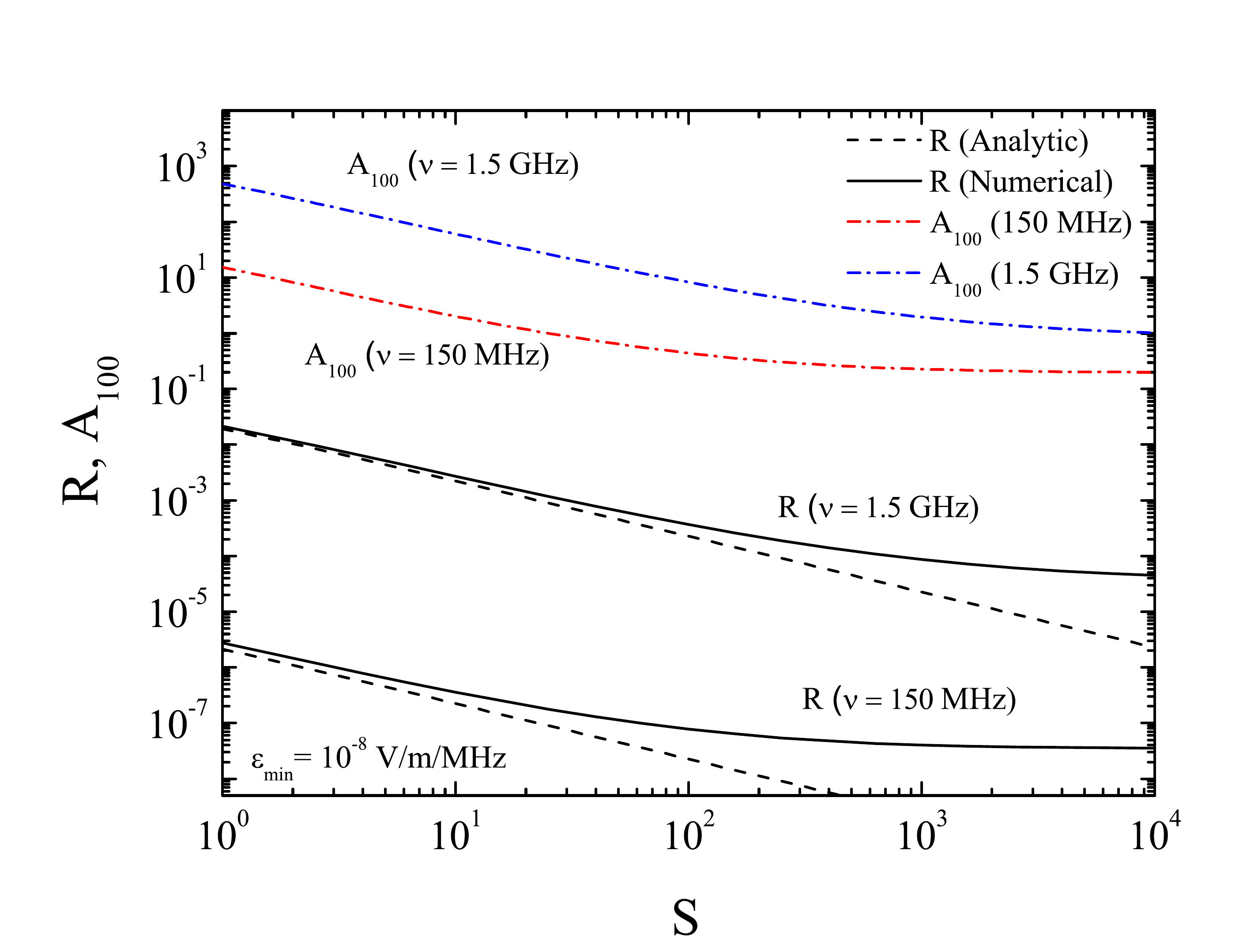}
\end{center}
\caption{For $E^2\Phi_\nu  = A\times 10^{-8}$ GeV/cm$^2$/s/sr, the ratio $R=\Gamma_{CR}/\Gamma_\nu$ for $A=1$ is shown with the solid lines where the full expression for the neutrino rate is used, while the dashed lines are evaluated using the ``neutrino approximation," with no attenuation for down-going neutrinos, as a function of $S=\sigma_{\nu N}/\sigma_{\nu N}^{SM}$ for $\varepsilon = 10^{-8} \, \rm V m^{-1} MHz^{-1}$. The dot-dashed lines show the minimum $A$ to produce one neutrino event in 100 hrs, as a function of $S$.
}
\label{fig:RAem8}
\end{figure}

\begin{figure}[h]
\begin{center}
\includegraphics[width=0.45\textwidth]{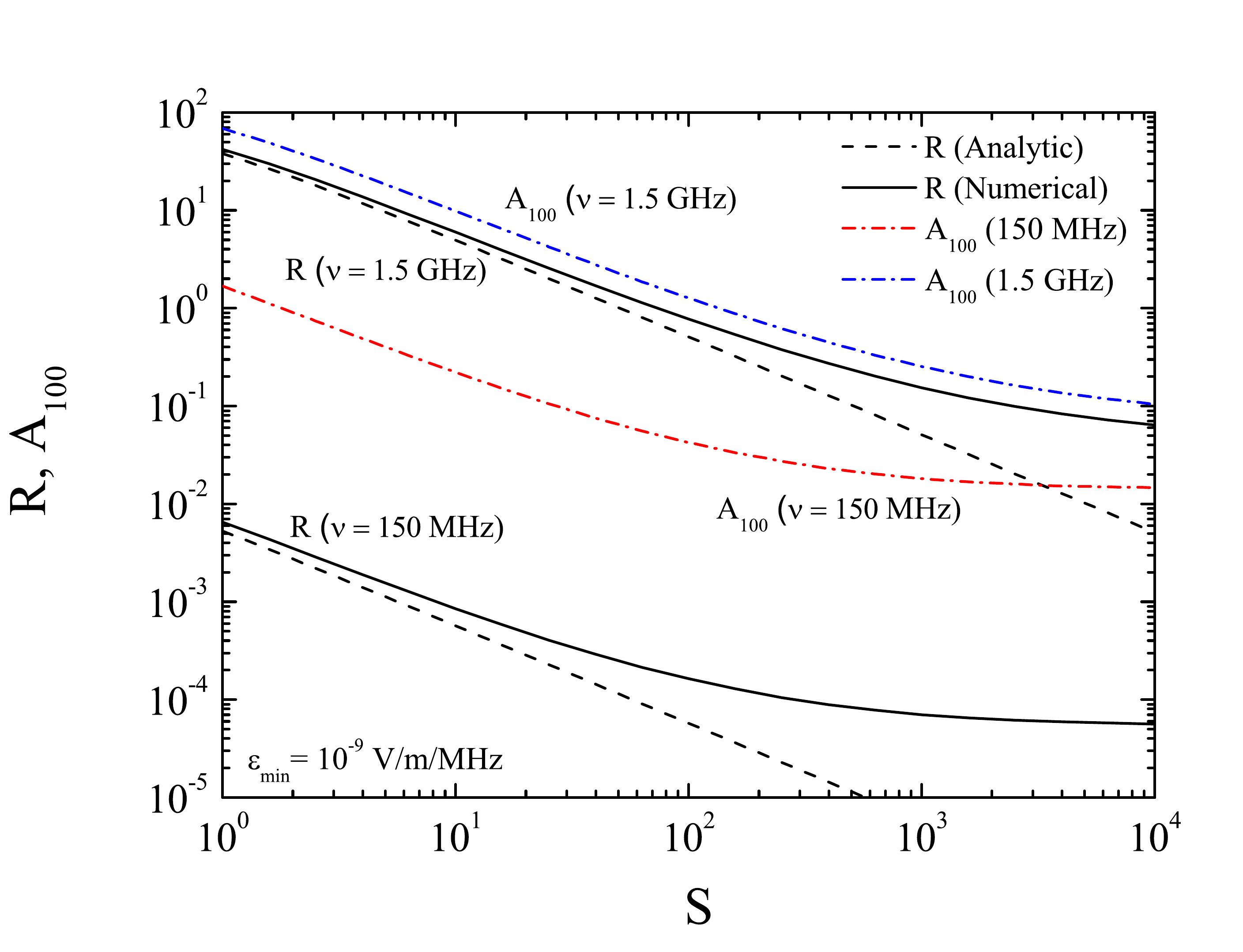}
\end{center}
\caption{The ratio $R$ and the minimum $A_{100}$, 
as in Fig. \ref{fig:RAem8} but 
for $\varepsilon = 10^{-9} \, {\rm V m^{-1} MHz^{-1}}$.}
\label{fig:RAem9}
\end{figure}

Taking a more generic approach, looking at the detectability
as a function of $S$, we can see the capabilities
of the lunar technique based on the GMJ formalism.
In order to study the sensitivity of radio Cherenkov detection 
to the neutrino flux, we take neutrino flux of the form, 
\be
E^2 \Phi_\nu(E_\nu) = A \times 10^{-8}\ {\rm GeV}/{\rm cm}^2/{\rm s/sr.}
\label{eq:em2}
\ee
We investigate the value of  $A$ such that when
 $A=A_{100}$, the theoretical prediction is one neutrino event for 100 hours of 
observation.  We consider not only the total event rates but also 
the ratio of the cosmic ray induced signal over the neutrino induced signal, 
\be
R=\frac{\Gamma_{CR}}{\Gamma_{\nu}}\ .  
\ee
In Figs. \ref{fig:RAem8} and \ref{fig:RAem9}, we show the two quantities, $R$ 
and the minimum $A=A_{100}$ required for one neutrino event in 100 hrs, for the
flux in eq. (\ref{eq:em2}) for $\varepsilon_{\rm min}=10^{-8}$ and $10^{-9}$ V/m/MHz,
respectively, as a function of $S=\sigma_{\nu N}/\sigma_{\nu N}^{SM}$
The solid lines in these figures show the ratio $R$ using the expression in eq. (A8) which produces the solid lines. This is labeled the 
``numerical" result. The dashed lines show the result using the ``neutrino approximation" in which 
attenuation of the downward neutrino flux is
not included.   The dot-dashed lines show the value of $A_{100}$ required to produce one
event in 100 hours.

For $\varepsilon = 10^{-8} \ \rm V/ m/ MHz$, for both frequencies shown (1.5 GHz and 150 MHz), $R\ll 1$. The deviation between the solid and dashed lines shows the expected result that eventually, the 
neutrino rate does not increase with cross section because of attenuation of the downward flux in the Moon. The dot-dashed lines showing the
minimum $A$ for a neutrino flux with an analytic form following eq. (\ref{eq:em2})
required for one neutrino event in 100 hrs also show the saturation effect due to attenuation.
Even with large neutrino cross sections ($S\gg 1$),
given 100 hrs, the minimum observable $A$ for $\nu=1.5$ GHz and $\varepsilon = 10^{-8} \, \rm V /m /MHz$ is $A_{100}\simeq 1$.
For values of $S$ closer to 1, the minimum observable $A$ is several hundred for 100 hrs of observation. 
This range of parameter space is already excluded by ANITA 
\cite{anita}, however, not in this energy range.  
The lower frequency of $\nu=150$ MHz probes a lower value of $A$,  from $A=1$ down to $A=0.2$. 

In Fig. \ref{fig:RAem9} we show the same ratio $R$ and 
$A_{100}$ for 
$\varepsilon = 10^{-9} \, {\rm V /m/ MHz}$.  In this 
case the values of $A_{100}$ needed to reach one 
neutrino event are much 
 lower than 
in the case when 
$\varepsilon = 10^{-8} \, {\rm V/ m/MHz}$.  This is due to the 
fact that lowering 
$\varepsilon$ we are probing neutrino fluxes 
at a lower energy where the flux is higher, 
as can be seen from Fig. 1.  However, for $\nu=1.5$ GHz, 
the ratio $R$ is also much larger, and for $S<10^2$, the CR background 
is larger than the neutrino signal.  For lower frequencies, when 
$\nu=150$ MHz, for any value of $S$, including $S=1$ which corresponds 
to the standard model neutrino cross section, the CR background is 
much smaller than the neutrino signal.  In general, lower frequencies 
give larger event rates, and operating at 
lower frequency has this advantage in detecting neutrino fluxes.  A theoretical
advantage at the lower frequencies is that the event rate relies less on the surface roughness
than for the higher frequencies. However, the observational challenge of atmospheric dispersion
of the signal at $\nu=150$ MHz is much more significant than at $\nu=1.5$ GHz. 

There are several radio telescopes that have recently looked for 
very high energy neutrinos using the Moon as a target. The 
RESUN Project used the extended very large array (EVLA) with 
$\nu\sim 1.5$ GHz and 
$\varepsilon_{\rm min}= 10^{-8}$ 
V/m/MHz 
\cite{Jaeger}.   
The Westebork Synthesis Radio Telescope 
(WSRT) operates in the frequency range of $115-180$ MHz and 
 has reported limit on 
neutrino flux of 
$E^2\Phi_\nu \sim 10^{-6} \ {\rm GeV/cm^2/s/sr}$ 
\cite{westerbork}.  
The Lunaska experiment at the Australia Telescope Compact Array 
(ATCA) \cite{lunaska} is operating in the region 
$\nu = 1.2 - 1.8$ GHz  and is 
expected to have sensitivities to lower neutrino fluxes than 
RESUN. At 
ultra-high energies ($E_\nu > 10^{14}$ GeV), NuMoon \cite{nuMoon} 
which uses 
the Westebork Synthesis radio Telescope in Netherlands, one of 
the most sensitive low-frequency experiment ($\nu =113 - 175$ MHz), 
will be able to place more stringent limit on the neutrino 
flux. 
LOFAR (Low Frequency Array), which covers 
frequencies between 
120 MHz and 240 MHz and between 10MHz and 80 MHz, is expected to 
lower their energy threshold down to $10^{11}$ GeV and with 
30 days of data taking will be probing 
the neutrino flux down to 
$E^2\Phi_\nu \sim  3 \times 10^{-9} \ {\rm GeV/cm^2/s/sr}$ 
assuming standard model neutrino interactions 
\cite{LOFAR}.  
Cosmic ray background for LOFAR is still sufficiently small 
relative to the neutrino signal.  The advantage of LOFAR over 
RESUN, for example, is that it operates at low frequency and 
has longer observation time.  However, it is important 
to note that lowering 
energy threshold down to $10^{10}$ GeV (i.e. lowering 
$\varepsilon_{min}$ below $10^{-10}$ 
V/m/MHz), 
would result in 
 larger neutrino rates but the cosmic ray background 
would become significant.  For neutrino fluxes lower than 
$E^2\Phi_\nu \sim  3 \times 10^{-9} \ {\rm GeV/cm^2/s/sr}$, 
LOFAR provides an excellent probe of physics beyond the 
Standard Model.

\acknowledgments
We would like to thank Ken Gayley, Robert Mutel and John Ralston.  I.S. would like to 
 thank Aspen Center for 
Physics for the hospitality.  
This work is supported in 
part by DOE contracts DE-FG02-91ER40664, 
DE-FG02-04ER41319 and 
DE-FG02-04ER41298.  

\appendix
\section{Effective aperture for downward particle fluxes}

Following Gayley, Jaeger and Mutel in Ref. \cite{gayley}, we evaluate the 
probability function $P(E)$ for downward incident particles with
\begin{eqnarray}
\nonumber
P(E)&=& \frac{1}{\pi}\frac{L_\gamma}{L_\nu} \int_{-\infty}^0 d\alpha \cos\alpha
\int_{-\infty }^{\infty} d\Delta \sin(\theta_c+\Delta)\\
&\times& \int_0^\infty d\phi\int_0^{z_{max}} dz e^{-\tau_\nu} {\cal H}_R
{\cal H}_D\xi\ ,
\label{eq:peappendix}
\end{eqnarray}
(eq. (12) of Ref. \cite{gayley}). For the sake of discussion, we will call the
incident particles neutrinos. Here, $\alpha$ is the angle the incident neutrino
makes with respect to the horizontal, with $\alpha<0$ indicating ``downward''
neutrinos. The quantity $\Delta$ characterizes the polar angle of the interior
ray solid angle relative to the cherenkov angle $\theta_c$ of the radio signal, and $\phi$ is its azimuthal angle. The factor $\xi$ accounts for the fact that the
rado signal encounters the lunar surface from the inside. The quantity $z$ is the
depth $h$ in the lunar regolith, normalized by $L_\gamma$. In this equation,
${\cal H}_R$ and ${\cal H}_D$ are there to satisfy the conditions that the radio ray
refracts (and does not totally internally reflect inside the moon), and that the rays are bright enough (detectable). Ref. \cite{gayley} has ${\cal H}_R$ written
in terms of two more integrals, accounting for surface roughness so that the surface
tilt
polar angle is $\sigma = \sigma_0 w$ and the azimuthal tilt direction is $\phi'$ relative to a smooth surface, with
\begin{equation}
{\cal H}_R = \frac{2}{\pi^{3/2}} \int_0^{\pi/2} d\phi'\int_{-\infty}^{\infty}
dw\ e^{-w^2} {\cal H}_{w,\phi'} \ .
\end{equation}
The Heaviside step function ${\cal H}_{w,phi'}$ enforces the requirement that the
ray emerge from inside the Moon. Finally, the factor $e^{-\tau_\nu}$ accounts
for the attenuation of the incident neutrino flux with
\begin{equation}
\tau_\nu = z\cdot L_\gamma/L_\nu\cdot 1 /|\sin\alpha| \ .
\end{equation}

The integral over $z$  is straightforward, with 
\begin{equation}
\label{eq:zint}
\int_0^{z_{max}}dz e^{-\tau_\nu} = \frac{|\sin \alpha| L_\nu}{L_\gamma}
\biggl( 1 - e^{-z_{max} L_\gamma/(|\sin\alpha | L_\nu)} 
\biggr) \ .
\end{equation}
The quantity $z_{max}$ is
\begin{equation}
\label{eq:zmax}
z_{max}\simeq \sin{\theta_c} f_0^2 \Biggl(
1-\frac{\Delta^2}{f_0^2\Delta_0^2}\Biggr)\ .
\end{equation}

The result in eq. (\ref{eq:zint}) is the origin of the two limits, the ``cosmic ray limit'' and the
``standard model neutrino limit." In the cosmic ray limit,
\begin{equation}
\label{eq:zcr}
\int_0^{z_{max}}dz e^{-\tau_\nu} \simeq \frac{|\sin \alpha| L_\nu}{L_\gamma}\quad
{\rm CR\ limit} \ ,
\end{equation}
since the argument of the exponential is a large negative number with $L_\gamma/L_\nu$ large. The neutrino limit involves the expansion of the
exponential in a power series, keeping the first non-zero term, so
\begin{equation}
\label{eq:znu}
\int_0^{z_{max}}dz e^{-\tau_\nu} \simeq z_{max}\quad {\rm neutrino\ limit} \ .
\end{equation}

For enhanced neutrino cross sections, where neither the cosmic ray nor neutrino limits
are applicable, we numerically integrate eq. (\ref{eq:peappendix}) in the
small angle limit, with the Heaviside functions enforced with downward
incident angles to give
\begin{eqnarray}
\nonumber
P(E)&\simeq & 
\frac{n_r\sin\theta_c}{\pi^{3/2}} \int_0^{\pi/2}d\phi '\Biggl[\int_{w_0}^{\infty}
dw\ \int_{-f_0\Delta_0 }^{f_0\Delta_0} d\Delta \\ \nonumber
&+& 
\int_{-w_0}^{w_0}
dw\ \int_{-w f_0 \Delta_0/w_0 }^{f_0 \Delta_0} d\Delta \Biggr]
\int_{\alpha_{min}}^0 d\alpha \\
&\times &
 e^{-w^2}|\alpha |\biggl( 1 - e^{-z_{max} L_\gamma/(|\alpha | L_\nu)} 
\biggr)
\end{eqnarray}
where $\alpha_{min} = -\Delta-w\sigma_0\cos\phi'=
-\Delta-w/w_0$.
Analytically, in the downward neutrino limit with $\sin\theta_0=\sqrt{n_r^2-1}/n_r$, this yields
$$P(E)= P_\nu (E) \simeq \frac{n_r - 1}{8 n_r} \frac{L_\gamma}{L_\nu}
f_0^3\Delta_0\bigl(f_0\Delta_0+ \frac{16}{3\pi^{3/2}}\sigma_0\Bigr)\ .
$$
In the cosmic ray limit, the integrals give

$$P(E) = P_{CR}(E)\simeq \frac{\sqrt{n_r^2-1}(f_0 \Delta_0)^3}{12}\Biggl(1+\frac{3}{4}\frac{\sigma_0^2}{f_0^2\Delta_0^2}
\Biggr)\ .$$

\end{document}